\documentclass[twocolumn,amsmath,amssymb,floatfix,eqsecnum,pra]{revtex4-1}
\usepackage{amsmath}
\usepackage{enumerate}
\usepackage{graphicx}
\begin{document}
\def \brho {{\rho} \hskip -4.5pt {\rho}}
\def \bnab {{\bf\nabla}\hskip-8.8pt{\bf\nabla}\hskip-9.1pt{\bf\nabla}}
\title{Application of axiomatic formal theory to the Abraham--Minkowski controversy}
\author{Michael E. Crenshaw}
\email{michael.e.crenshaw4.civ@mail.mil}
\affiliation{Charles M. Bowden Research Laboratory, US Army Combat Capabilities Development Command (DEVCOM) - Aviation and Missile Center, Redstone Arsenal, AL 35898, USA}
\begin{abstract}
A transparent linear dielectric in free space that is illuminated by a
finite quasimonochromatic field is a thermodynamically closed system.
The energy--momentum tensor that is derived from Maxwellian continuum
electrodynamics for this closed system is inconsistent with conservation of
momentum; that is the foundational issue of the century-old
Abraham--Minkowski controversy.
The long-standing resolution of the Abraham--Minkowski controversy is to
view continuum electrodynamics as a subsystem
and write the total energy--momentum tensor as the sum of an
electromagnetic energy--momentum tensor and a phenomenological material
energy--momentum tensor.
We prove that if we add a material energy--momentum tensor to the 
electromagnetic energy--momentum tensor, in the prescribed manner, then the
total energy, the total linear momentum, and the total angular momentum are
constant in time, but other aspects of the conservation laws are violated.
Specifically, the four-divergence of the total (electromagnetic plus
material) energy--momentum tensor is self-inconsistent and violates
Poynting's theorem.
Then, the widely accepted resolution of the Abraham--Minkowski dilemma is
proven to be manifestly false.
The fundamental physical principles of electrodynamics, conservation laws,
and special relativity are intrinsic to the vacuum; the extant dielectric
versions of physical principles are non-fundamental extrapolations from the
vacuum theory.
Because the resolution of the Abraham--Minkowski momentum controversy is
proven false, the extant theoretical treatments of macroscopic
continuum electrodynamics, dielectric special relativity, and
energy--momentum conservation in a simple linear dielectric are
mutually inconsistent for macroscopic fields in matter.
We derive mutually consistent theoretical treatments of physical processes
in an isotropic, homogeneous, linear dielectric-filled,
flat, non-Minkowski, continuous material spacetime.
The Abraham--Minkowski momentum controversy has a robust resolution in the
new Lagrangian field theory of continuum electrodynamics.
\hfill
\end{abstract}
\date{\today}
\maketitle
\par
\section{Introduction}
\par
\subsection{Physical Setting}
\par
The foundations of the energy--momentum tensor and the associated
tensor-based spacetime conservation theory come to electrodynamics from
classical continuum dynamics where the divergence theorem is applied to
a Taylor series expansion of the density field of an intrinsic
property (\textit{e.g.,} mass, particle number) of an unimpeded,
inviscid, incoherent flow of non-interacting particles (molecules, dust,
etc.) in the continuum limit (fluid or particle number field, for example)
in an otherwise empty volume \cite{BIFox}.
Although the continuous formulation of conservation principles was
originally the provenance of fluid mechanics (continuum dynamics), the
energy and momentum conservation properties of light propagating in the
vacuum were long-ago cast in the energy--momentum tensor formalism in which
the light field (flow of non-interacting photons in the
continuum limit) plays the role of the continuous fluid \cite{BILL}.
However, extending the tensor-based theory of energy--momentum
conservation of the continuous light field in the vacuum to propagation
of the light field in a linear dielectric medium, also in the
continuum limit, has proven to be persistently problematic, as exemplified
by the more-than-century-old Abraham--Minkowski momentum
controversy \cite{BIMin,BIAbr,BIRL,BIBrev,BIAMC3,BIKemplatest,BIPfei,BIAMC2,BIAMC4,BIAMC5, BIObuk,BIObuk2,BIMuka,BIMuka2,BIBarn,BIPeierls,BIBalazs,BIGord,BIMol,BIBahder,BIJMP,BISPIE,BIBrev2,BINewExp1, BINewExp2,BISheMan,BISheBrev,BIBethune,BIBrevnew,BIBrevnew2}.
\par
The origin story of the Abraham-Minkowski controversy is that the Minkowski
energy--momentum tensor is not diagonally symmetric.
Motivated by the need to preserve the principle of conservation of angular
momentum, Abraham symmetrized the energy--momentum tensor by an ad hoc
redefinition of the linear momentum density to be the Poynting energy flux
vector divided by $c^2$.
The issue of the lack of symmetry in the Minkowski energy--momentum tensor
has since been overtaken by substantive problems with conservation of
linear momentum for both the Minkowski and Abraham energy--momentum
tensors.
The modern resolution of the Abraham--Minkowski momentum controversy is
to decide that the electromagnetic energy, the electromagnetic angular
momentum, and the electromagnetic linear momentum are features of an
electromagnetic subsystem.
The addition of a phenomenological material subsystem completes the total
system.
Although the ``modern'' resolution \cite{BIGord,BIPenHaus} has been around 
for 50 years, or so, it still isn't working out quite right.
\par
A gradient-index antireflection-coated right rectangular block of a simple
(absorption-negligible, isotropic, homogeneous, lowest-order dispersive)
linear dielectric material, with refractive index $n$, situated in
free-space that is illuminated at normal incidence by a finite
quasimonochromatic field with center frequency $\omega_p$ is a
\textit{thermodynamically closed system} that consists of the field, the
dielectric material, and any other pertinent subsystems.
Integrating over all-space $\Sigma$,
the Abraham (linear) momentum
\begin{equation}
{\bf G}_A=\int_{\Sigma} \frac{{\bf E}\times{\bf B}}{c} dv 
\label{EQw1.01}
\end{equation}
and the Minkowski (linear) momentum
\begin{equation}
{\bf G}_M=\int_{\Sigma} \frac{n^2{\bf E}\times{\bf B}}{c} dv 
\label{EQw1.02}
\end{equation}
are not constant in time as a finite electromagnetic pulse enters
the dielectric from the
vacuum \cite{BIBrev,BIAMC3,BIKemplatest,BIPfei,BIAMC2}.
The fact that the electromagnetic momentum is not conserved means that 
a portion of the incident electromagnetic momentum is being transferred to
some other subsystem and the dielectric material is the only other
identifiable component of the system.
\par
When the field is entirely within an antireflection-coated block of a 
homogeneous simple linear medium, the Abraham momentum is less than the
incident (from vacuum) momentum by a nominal factor of $n$.
The Minkowski momentum is greater than the Abraham momentum by a factor
of $n^2$ and it is therefore greater than the (vacuum) incident momentum
by a factor of $n$.
The fact that neither the Abraham momentum nor the Minkowski momentum is
globally conserved is well-known in the scientific
literature \cite{BIBrev}.
Nevertheless, many researchers \cite{BIPfei,BIObuk}
\textit{justifiably} claim that the Minkowski (linear) momentum is
conserved or ``almost'' conserved based on the four-divergence of the
Minkowski energy--momentum tensor being negligible, which corresponds to a
local conservation law \cite{BIGiu}, $\partial_{\beta}{\sf T}^{i\beta}=0$,
see Eq.~(\ref{EQw3.07}).
But, a factor of $n$ is not a perturbation and any
claim \cite{BIPfei,BIObuk} that the Minkowski
momentum (or the Minkowski tensor) is conserved or ``almost'' conserved is
manifestly \textit{false} based on the global conservation
condition \cite{BIGiu,BILL}, $P^{i}(t)=P^{i}(t_0)$,
see Eq.~(\ref{EQw3.08}).
Pfeifer, Nieminen, Heckenberg, and Rubinsztein-Dunlop \cite{BIPfei}
present both facts from the extant literature in their comprehensive review
article showing that the Minkowski momentum is greater than the incident
momentum by a factor of $n$ in their Sec.~9A and noting that the Minkowski
momentum is almost conserved in their Sec.~8C.
This unremarked contradiction between local and global conservation laws
\cite{BIGiu}, appears to be one of many factors that contribute to the
extraordinary longevity of the Abraham--Minkowski controversy.
\par
In time gone by, the necessity for a medium in which light waves propagate
proved the existence of the ether.
The properties of the ether, including ether drag and partial ether drag,
were determined from the observed characteristics of light propagation in
the vacuum.
Likewise, the necessity for global conservation of linear momentum in a 
thermodynamically closed field-plus-matter system proves the existence of
a material momentum that is associated with the movement of matter in a
microscopic picture of a dielectric
substructure \cite{BIAMC3,BIKemplatest,BIPfei,BIAMC2}.
We note that such a microscopic subsystem is adverse to the continuum
limit that is a foundational concept of continuum electrodynamics.
Nevertheless, the assumption of identifiable electromagnetic and
material component subsystem momentums is commonly made and assumed to
be settled physics in the scientific literature 
\cite{BIAMC3,BIKemplatest,BIPfei,BIAMC2}.
However, the nature of the material subsystem is elusive and controversial.
\par
While the physical origin and characteristics of the phenomenological
material momentum have been debated for many decades, experimental
demonstrations of a material momentum have been variously inconsistent,
contradictory, and refuted
\cite{BIBrev,BIBrev2,BINewExp1,BINewExp2,BISheMan,BISheBrev,BIBethune}.
Likewise, theoretical treatments of the Abraham--Minkowski controversy
\cite{BIAMC3,BIKemplatest,BIPfei,BIAMC2,BIAMC4,BIAMC5,BIObuk,BIObuk2,BIMuka,BIMuka2,BIBarn,BIPeierls,BIBalazs,BIGord,BIMol,BIBahder,BIJMP,BISPIE}
have not provided a consistent physical solution.
In 2007,
Pfeifer, Nieminen, Heckenberg, and Rubinsztein-Dunlop \cite{BIPfei}
reviewed the state of the field and concluded that the issue had been
settled for some time:
``When the appropriate accompanying energy--momentum tensor for the
material medium is also considered, experimental predictions of the
various proposed tensors will always be the same, and the preferred form
is therefore effectively a matter of personal choice.''
Three years later, Barnett \cite{BIBarn} offered a more restrictive
resolution of the Abraham--Minkowski debate by contending that the total
momentum ${\bf G}_{total}$ for the medium and the field is composed of
either the Minkowski canonical momentum ${\bf G}_{M}$ or the Abraham
kinetic momentum ${\bf G}_{A}$ supplemented by the appropriate canonical
material momentum ${\bf G}_{can}$ or kinetic material
momentum ${\bf G}_{kin}$.
Although the Barnett \cite{BIBarn} and
Pfeifer, Nieminen, Henckenberg, and Rubinsztein-Dunlop \cite{BIPfei}
theories are based on fundamental principles, Barnett's restriction to two
simultaneous physically motivated electromagnetic momentums is
\textit{contradicted} by the mathematical tautology that underlies the
analysis of Pfeifer, et al., and vice-versa.
\par
In order to present a concrete example for discussion, we use the common
Barnett \cite{BIBarn} model for the field and matter components of the
total energy--momentum tensor.
The material momentums are constructed such that the total
(electromagnetic plus material) linear momentum \cite{BIAMC2,BIBarn}
\begin{subequations}
\begin{equation}
{\bf G}_{total}={\bf G}_{incident}={\bf G}_{em}+{\bf G}_{material}
\label{EQw1.03a}
\end{equation}
\begin{equation}
{\bf G}_{total}={\bf G}_{A}+{\bf G}_{kin}={\bf G}_{M}+{\bf G}_{can}
\label{EQw1.03b}
\end{equation}
\label{EQw1.03}
\end{subequations}
of a finite pulse of the continuous light field in a continuous simple
linear dielectric material with no other consequential subsystems is
constant in time as required by global conservation
\cite{BIPfei,BIFox,BILL,BIGiu} for a continuous fluid.
Clearly, ${\bf G}_{kin}=(1-1/n){\bf G}_{incident}$ and
${\bf G}_{can}=(1-n){\bf G}_{incident}$ when the electromagnetic pulse
is fully within an antireflection-coated, isotropic, homogeneous, 
transparent linear dielectric medium.
The material energy is constructed such that the total (electromagnetic
plus material) energy is equal to the incident energy \cite{BIPfei}
\begin{subequations}
\begin{equation}
U_{total}=U_{incident}= U_{em}+U_{material} 
\label{EQw1.04a}
\end{equation}
\begin{equation}
U_{total}= U_{A}+U_{kin}= U_{M}+U_{can} 
\label{EQw1.04b}
\end{equation}
\label{EQw1.04}
\end{subequations}
and is constant in time as the light pulse propagates from the vacuum
into the antireflection-coated material.
The total energy--momentum tensor is the sum of the Abraham
energy--momentum tensor and a kinetic material energy--momentum tensor.
In addition, the total energy--momentum tensor is the sum of the
Minkowski energy--momentum tensor and a canonical material
energy--momentum tensor.
That is \cite{BIPfei}, 
\begin{subequations}
\begin{equation}
{\sf T}_{total}^{\alpha\beta}= {\sf T}_{incident}^{\alpha\beta}=
{\sf T}_{em}^{\alpha\beta}+ {\sf T}_{material}^{\alpha\beta}
\label{EQw1.05a}
\end{equation}
\begin{equation}
{\sf T}_{total}^{\alpha\beta}=
{\sf T}_{A}^{\alpha\beta}+{\sf T}_{kin}^{\alpha\beta}
={\sf T}_{M}^{\alpha\beta}+{\sf T}_{can}^{\alpha\beta} \, .
\label{EQw1.05b}
\end{equation}
\label{EQw1.05}
\end{subequations}
Obviously, Brevik's \cite{BIBrevnew} admonitions against the application of
conservation laws to subsystems do not apply to conservation of the total
energy, the total linear momentum, the total angular momentum, or the total
energy--momentum tensor of the \textit{complete} and \textit{closed}
field-plus-matter system that is considered here and in
Refs.~\cite{BISPIE,BIBahder,BIJMP}.
\par
\par
According to the Scientific Method, a scientific hypothesis must
result in a unique testable prediction of physical quantities.
There are many examples in the experimental record in which the
interpretation of momentum experiments is unrestricted with experiments
that prove the Minkowski electromagnetic momentum later being shown
to prove the Abraham electromagnetic momentum and
vice versa \cite{BIBrev}.
Likewise, experiments that disprove the Minkowski momentum are later
re-analyzed to confirm the Minkowski momentum and similarly for
the Abraham momentum \cite{BIBrev}.
The non-uniqueness of the electromagnetic momentum and the 
non-uniqueness of the material momentum for light in a dielectric are
contrary to Popper's criterion of falsifiability and constitute a
violation of the Scientific Method.
\par
The prior work appears to present a very complex situation because some
experiments support the Abraham definition of momentum and other
experiments support the Minkowski momentum formula.
This might suggest that the Scientific Method needs to be malleable in
order to accommodate the experimentally proven non-unique incommensurate
momentums.
However, appeal to complexity is a fallacious application of the
Scientific Method.
The problem is that electromagnetic momentum is not measured directly.
Instead the force due to optical pressure on a mirror inserted in a
fluid dielectric is measured and related to the change in
electromagnetic momentum \cite{BIBrev,BIPfei,BIExp}.
Using the Abraham, Minkowski, or other momentum formula to relate the
measured quantity (force) to the momentum creates a circular, self-proving
theory, see Sec.~7. 
\par
We are accustomed to having well-designed experiments either prove
or disprove scientific hypotheses.
However, \textit{experiments cannot, in either principle or practice,
provide a definitive resolution of the Abraham--Minkowski
controversy} \cite{BIBrevnew} because the Scientific Method has been
abrogated by the subsystem separation.
In order to be theoretically definitive, \textit{the present work is
based almost entirely on the unique, unseparated, globally conserved total
(field plus matter plus other) momentum or the corresponding total momentum
density} of a thermodynamically closed system with appropriate system
boundaries.
\par
In a thermodynamically closed system, the total momentum, like the total
energy, is a known quantity that is uniquely determined from the initial
conditions by being constant in time.
Then the total energy density and the total linear momentum density
components of the total energy--momentum tensor are uniquely decided by
time independence of the spatially integrated total energy and total
momentum densities (global conservation law) \cite{BISPIE,BIBahder,BIJMP}.
The construction of the unique total (electromagnetic plus material plus 
other) energy--momentum tensor \cite{BISPIE,BIBahder,BIJMP} from these
total energy density and total momentum density components gets us
congruent with the Scientific Method and definitively resolves the
Abraham--Minkowski controversy.
That is the generally accepted resolution of the Abraham--Minkowski
controversy.
{\bf Except that there are problems here, too.}
Applying the four-divergence operator to the unique total
energy--momentum tensor, one obtains spacetime conservation laws in the
form of continuity equations for the total energy and the total linear
momentum \cite{BIPfei,BISPIE,BIBahder,BIJMP}.
It is easily argued, on physical grounds, that the conservation law for
the total energy that is obtained by applying the four-divergence operator
to the total energy--momentum tensor cannot be incommensurate with the
Poynting theorem that is systematically derived from the Maxwell--Minkowski
field equations for a simple linear dielectric.
It is also \textit{easily proven} mathematically that
{\it i}) the energy conservation law, derived from the four-divergence
of the total energy--momentum tensor, is self-inconsistent because its
two non-zero terms depend on different powers of the refractive index
and {\it ii}) the conservation law for the total energy that is derived
from the four-divergence of the total energy--momentum tensor is
undeniably incommensurate with the Poynting theorem \cite{BIJMP}.
The contradiction of sound physical arguments by mathematical reality
seems to also contribute to the longevity of the dispute.
\par
In summary, we have ``fixed'' the Abraham--Minkowski momentum conservation
problem \textit{in the prescribed manner} \cite{BIPfei,BIBarn} only
to have contradictions appear in a different form elsewhere.
The current author \cite{BISPIE,BIBahder,BIJMP} made the
\textit {ansatz} of a Ravndal \cite{BIFinn} refractive index-dependent
\textit{material} four-divergence operator and demonstrated consistency
between the field equations and total energy--momentum conservation
laws, at the cost of apparent problems with special relativity and the
Fresnel relations.
Moveable contradictions with multiple resolutions are characteristic of an
inconsistent theoretical foundation with physically motivated ad hoc
patches.
A comprehensive approach to the basic theories and the relations between
field theory, conservation laws, special relativity, spacetime, and
boundary conditions for a quasimonochromatic field propagating in a simple
linear dielectric medium is absolutely required.
\par
\par
\subsection{Procedure}
\par
Classical continuum electrodynamics can be treated mathematically as a
formal system in which the macroscopic Maxwell--Minkowski equations and
the constitutive relations are the axioms.
Theorems are derived from the axioms using algebra and calculus.
In this article, we derive theorems of the formal theory of
Maxwell--Minkowski continuum electrodynamics from explicitly stated axioms
using substitutions of explicitly defined quantities.
Steps are kept small so that all readers should be quite satisfied that
there are no implicit axioms and no manifest deficiencies in the
derivations.
Noting that spacetime conservation laws \cite{BIFox,BILL,BIPfei,BIGiu}
(that are reviewed in Sec.~3) and the macroscopic Maxwell field equations
for a linear dielectric medium are distinct laws of physics, we show that
valid theorems of Maxwellian continuum electrodynamics are proven false
by the spacetime conservation laws for an inviscid, incoherent flow
of non-interacting particles (photons) in the continuum limit (light 
field) through the continuous dielectric medium and in the absence of
external forces, pressures, constraints, or other impediments (unimpeded
flow).
\par
When a valid theorem of an axiomatic formal theory is proven false by an
accepted standard then it is proven that one or more axioms of the
formal theory are false.
Alternatively, the standard with which the theorem is being compared is
false or both the theory and the standard can be simultaneously false.
Therefore, the extant theoretical treatments of spacetime energy--momentum
conservation and Maxwellian continuum electrodynamics \textit{must} be
regarded as being mutually inconsistent in a region of space in which the
effective speed of light is $c/n$.
\par
Nevertheless, there is a generally accepted ``fix'' in which a physically
motivated material momentum and a physically motivated material
energy--momentum tensor are added to the rigorously derived Maxwellian
electromagnetic quantities \cite{BIPfei,BIBarn}.
The total field-plus-matter linear momentum is constant in time so that
the modified momentum and energy--momentum tensor are proven true by global
conservation principles.
Then, we prove that the physically motivated fix is \textit{false} because
the four-divergence of the total (electromagnetic plus material)
energy--momentum tensor is self-inconsistent, violates Poynting's theorem,
and violates spacetime conservation laws.
\par
Obviously, we cannot attach any credence to a contradiction between
fundamental physical laws (Maxwell's equations and spacetime conservation
laws) because the contradiction would have been discovered by now; unless
such a contradiction was found but not recognized.
For over 50 years, global conservation principles have been used to justify
a phenomenological material momentum that remediates the contradiction
between the tensor energy--momentum continuity equation and the global
conservation law.
Assuming that fixing the problem results in a correct theory, most
researchers have failed to notice that the corrected continuity equation
now violates the local conservation law.
The current author \cite{BIJMP} noticed and further patched the
energy--momentum theory to make the continuity equation consistent with
both the local and the global conservation laws \cite{BIGiu}, but at the
expense of a contradiction with Einstein--Laue special relativity in a
dielectric.
The patches to the theory simply move the contradictions around as is
characteristic of an underlying theory that is self-inconsistent.
\par
Having proven the original version and the phenomenologically patched
version of Maxwellian continuum electrodynamics to be manifestly false in a
simple linear dielectric medium, it is customary to propose an alternative
theory.
In the continuum limit, the electromagnetic field and the dielectric medium
are continuous at all length scales.
We define an isotropic, homogenous, linear dielectric-filled, flat,
non-Minkowski, continuous material spacetime $S_d(\bar x_0,x,y,z)$
and we derive a new non-Maxwellian continuum electrodynamics from
Lagrangian field theory in this non-Minkowski spacetime.
\par
The basis functions of Maxwellian electrodynamics in Minkowski spacetime
$S_v(x_0,x,y,z)$ are 
$[\exp({-i((\omega/c)x_0-(n\omega/c){\bf \hat k}_0\cdot{\bf r})})+c.c.]/2$.
Here, $x_0=ct$ is the time-like coordinate of Minkowski spacetime and 
${\bf\hat k}_0$ is a unit vector in the direction of propagation.
In the non-Minkowski spacetime $S_d(\bar x_0,x,y,z)$, the basis
functions of propagating electromagnetic fields,
$[\exp(-i((n\omega/c)\bar x_0-(n\omega/c){\bf\hat k}_0\cdot{\bf r}))
+c.c.]/2$, are what we would expect for monochromatic, or
quasimonochromatic, fields propagating at speed $c/n$ in a linear
dielectric.
Consequently, there is a fundamental difference in approach between an
organically continuum-based non-perturbative electrodynamics and the
macroscopically averaged effects of microscopic fields interacting
perturbatively with a collection of individual particles in the vacuum.
Furthermore, the assumptions, approximations, limits, and averages that are
implicit in the macroscopic model are not reversible.
We cannot re-discretize or un-average a continuous
dielectric with macroscopic refractive index $n$ any more than we can
ascertain the velocity of a particle in an ideal gas with
temperature $T$.
\par
We apply Lagrangian field theory in the isotropic, homogeneous,
linear dielectric-filled, flat, non-Minkowski, continuous material
spacetime $S_d(\bar x_0,x,y,z)$ to systematically derive
equations of motion for macroscopic fields in a simple linear
dielectric.
\textit{These equations of motion are the axioms of a new formal theory
of continuum electrodynamics.}
Although the new continuum electrodynamics and Maxwellian continuum
electrodynamics are disjoint, there is sufficient commonality between the
new equations of motion and the macroscopic Maxwell--Minkowski field
equations in a dielectric that the extensive theoretical and experimental
work that is nearly correctly described by the macroscopic
Maxwell--Minkowski equations has an equivalent formulation in the new
theory.
More interesting is the work that we can do with the new formalism of
continuum electrodynamics that is improperly posed in the standard
Maxwell theory of continuum electrodynamics.
These cases will typically involve the invariance or tensor properties
of the set of coupled equations of motion for the macroscopic fields.
This interpretation is borne out in our common experience: the
macroscopic Maxwell--Minkowski equations produce exceedingly accurate
experimentally verified predictions of simple phenomena, 
but fail to render a unique, uncontroversial, verifiable prediction of
energy--momentum conservation for electromagnetic fields propagating in
a simple linear dielectric medium.
An analogy can be made to Newtonian dynamics that accurately described
all known dynamical phenomena until confronted with Lorentz length
contraction and time dilation, the Michelson--Morley \cite{BIMichMor}
experiment, and Einstein's relativity.
\par
Spacetime conservation laws are distinct from the energy-like and
momentum-like evolution theorems that are systematically derived from
the macroscopic Maxwell--Minkowski field equations, although the evolution
theorems are sometimes incorrectly referred to as the macroscopic
electromagnetic energy and momentum conservation laws.
In continuum dynamics, spacetime conservation laws are derived for the
case of an unimpeded (no external forces, pressures, or constraints),
inviscid, incoherent flow of non-interacting particles (dust, molecules of
a fluid, etc.) in the continuum limit in an otherwise empty
volume \cite{BIFox}.
Applying the divergence theorem to a Taylor series expansion of a
density field of a conserved property (mass, particle number,
etc.) in an empty Minkowski spacetime results in a continuity equation 
(conservation law) for the conserved property \cite{BIFox}.
We show that applying the same derivation procedure to a non-empty,
linear dielectric-filled, isotropic, homogeneous, flat, non-Minkowski,
continuous material spacetime $S_d(\bar x_0,x,y,z)$ produces a continuity
equation for a conserved property in a simple linear dielectric that
requires differentiation with respect to the independent time-like variable
$\bar x_0$ instead of $x_0$.
\par
We construct the diagonally symmetric, traceless, total
energy--momentum tensor as an element of
a valid theorem of the new formal theory of continuum electrodynamics.
We show that the spatial integrals of the total energy density and
the total momentum density are constant in time as the field propagates
from the vacuum into the medium and that theorems of the new
formal theory correspond to continuity equations for the conserved
properties in a dielectric-filled spacetime.
The tensor properties of the new formulation of continuum
electrodynamics constitute a definitive resolution of the
Abraham--Minkowski dilemma.
\par
The \textit{unique} total energy--momentum tensor that is derived by the
new formal theory is \textit{entirely electromagnetic in nature}.
Consequently, there is no need to assume any ``splitting'' of the
total energy, the total momentum, or the total angular momentum into
field and matter subsystems in order to satisfy the spacetime
conservation laws for a simple linear dielectric:
{\it i}) In the case of quasimonochromatic optical radiation incident on
a stationary homogeneous simple linear dielectric draped with a
gradient-index antireflection coating, the surface forces are
negligible. Then, the total energy, the total linear momentum, and the
total angular momentum are purely electromagnetic and the dielectric
remains internally and externally stationary.
{\it ii}) In the absence of an antireflection coating, the dielectric
block, as a whole, acquires (material) momentum due to the optically
induced surface pressure that is associated with Fresnel reflection and
this must be treated by boundary conditions, not by a hypothetical,
unobservable, internal material momentum.
The dielectric remains internally stationary.
\par
In this work, the index convention for Greek letters is that they
belong to $\{0,1,2,3\}$ and lower case Roman indices from the middle
of the alphabet are in $\{1,2,3\}$.
Coordinates $(x_1,x_2,x_3)$ correspond to $(x,y,z)$, as usual.
The Einstein summation convention in which repeated indices are summed
over is employed.
\par
\section{Maxwellian Continuum Electrodynamics}
\par
There are several representations of the macroscopic Maxwell equations.
Alternative formulations that are associated with Amp\`ere,
Chu, Lorentz, Minkowski, Peierls, and
others \cite{BIPenHaus,BIAMC4,BIKemplatest,BIPeierls,BIMol,BIJMP}
are sometimes used to emphasize various features of classical
electrodynamics in ponderable matter.
\par
In order to present a familiar and concrete narrative, we start with the
common Minkowski representation of the macroscopic Maxwell field
equations \cite{BIMar,BIGriff,BIJackson,BIZangwill}
\begin{subequations}
\begin{equation}
\nabla\times{\bf E}+\frac{1}{c}\frac{\partial{\bf B}}{\partial t}=0
\label{EQw2.01a}
\end{equation}
\begin{equation}
\nabla\times{\bf H}-\frac{1}{c}\frac{\partial{\bf D}}{\partial t}=
\frac{{\bf J}_f}{c}
\label{EQw2.01b}
\end{equation}
\begin{equation}
\nabla\cdot{\bf B}=0
\label{EQw2.01c}
\end{equation}
\begin{equation}
\nabla\cdot{\bf D}= \rho_f \, .
\label{EQw2.01d}
\end{equation}
\label{EQw2.01}
\end{subequations}
The macroscopic Minkowski fields,
${\bf E}({\bf r},t)$, ${\bf D}({\bf r},t)$,
${\bf B}({\bf r},t)$, and ${\bf H}({\bf r},t)$, are functions of position
${\bf r}$ and time $t$.
Here, ${\bf J}_f({\bf r},t)$ is the free charge current density and
$\rho_f({\bf r},t)$ is the free charge density.
Equations (\ref{EQw2.01}) are the axioms of Maxwell--Minkowski
electrodynamics for macroscopic fields in matter.
\par
The axioms, Eqs. (\ref{EQw2.01}), can be operated upon using standard
algebra and calculus to derive theorems.
For example, substituting the temporal derivative of the Gauss law,
Eq.~(\ref{EQw2.01d}), into the divergence of the Maxwell--Amp\`ere law,
Eq.~(\ref{EQw2.01b}), we obtain the continuity equation
\begin{equation}
\frac{\partial \rho_f}{\partial t} + \nabla\cdot {\bf J}_f =0 
\label{EQw2.02}
\end{equation}
for free charges.
\par
We take the scalar product of Eq.~(\ref{EQw2.01a}) with ${\bf H}$ and
the scalar product of (\ref{EQw2.01b}) with ${\bf E}$ and subtract the
results to produce a continuity equation
\begin{equation}
\frac{1}{c}\left (
{\bf E}\cdot\frac{\partial{\bf D}}{\partial t}
+{\bf H}\cdot \frac{\partial{\bf B}}{\partial t}
\right )
+\nabla\cdot ({\bf E}\times{\bf H})=-\frac{{\bf J}_f}{c}\cdot{\bf E} 
\label{EQw2.03}
\end{equation}
that is also a theorem of the formal theory of macroscopic
Maxwell--Minkowski continuum electrodynamics.
\par
Adding
the vector product of ${\bf B}$ with Eq.~(\ref{EQw2.01b}),
the vector product of ${\bf D}$ with Eq.~(\ref{EQw2.01a}),
the product of Eq.~(\ref{EQw2.01c}) with $-{\bf H}$, and
the product of Eq.~(\ref{EQw2.01d}) with $-{\bf E}$
produces
$$
\frac{1}{c}\frac{\partial}{\partial t}({\bf D}\times{\bf B})
+{\bf D}\times(\nabla\times {\bf E})
+{\bf B}\times(\nabla\times{\bf H})
$$
\begin{equation}
-(\nabla\cdot{\bf D}){\bf E} - (\nabla\cdot{\bf B}){\bf H} =
-\rho_f{\bf E} -\frac{1}{c}{\bf J}_f\times{\bf B}
\label{EQw2.04}
\end{equation}
that is also a valid theorem of Maxwellian continuum electrodynamics.
\par
In the absence of charges and charge currents, the energy and momentum
continuity equations, Eqs.~(\ref{EQw2.03}) and (\ref{EQw2.04}), are
\begin{subequations}
\begin{equation}
\frac{1}{c}\left (
{\bf E}\cdot\frac{\partial{\bf D}}{\partial t}
+{\bf H}\cdot \frac{\partial{\bf B}}{\partial t}
\right )
+\nabla\cdot ( {\bf E}\times{\bf H})= 0 
\label{EQw2.05a}
\end{equation}
$$
\frac{1}{c}\frac{\partial}{\partial t}({\bf D}\times{\bf B})
+{\bf D}\times(\nabla\times {\bf E})
+{\bf B}\times(\nabla\times{\bf H})
$$
\begin{equation}
-(\nabla\cdot{\bf D}){\bf E} - (\nabla\cdot{\bf B}){\bf H} = 0\, .
\label{EQw2.05b}
\end{equation}
\label{EQw2.05}
\end{subequations}
\par
In order to examine the importance of charges and charge currents
to the existing momentum conservation theory, let us
consider a different derivation of Eq.~(\ref{EQw2.05b}).
The momentum density ${\bf p}$ imparted to free charges can be
calculated by postulating the Lorentz force density 
\cite{BIMar,BIGriff,BIJackson,BIZangwill}
\begin{equation}
\frac{d{\bf p}}{dt}={\bf f}_L=
\rho_f{\bf E}+ \frac{{\bf J}_f}{c}\times{\bf B} 
\label{EQw2.06}
\end{equation}
as a physical law.
Now, eliminate the sources in favor of the fields using the Gauss law,
Eq.~(\ref{EQw2.01d}), to eliminate $\rho_f$ and using Faraday's law, 
Eq.~(\ref{EQw2.01b}),
to eliminate ${\bf J}_f$.
Then the momentum density ${\bf p}$ imparted to the free charges can be
calculated as \cite{BIMar,BIGriff,BIJackson,BIZangwill}
\begin{equation}
\rho_f{\bf E}+ \frac{{\bf J}_f}{c}\times{\bf B} =
(\nabla\cdot {\bf D}){\bf E}+ \left (\nabla\times{\bf H}
-\frac{1}{c}\frac{\partial{\bf D}}{\partial t}\right )\times{\bf B} \, .
\label{EQw2.07}
\end{equation}
Substituting the calculus identity
\begin{equation}
\frac{\partial}{\partial t}({\bf D}\times{\bf B}) =
\frac{\partial{\bf D}}{\partial t}\times{\bf B} +
{\bf D}\times \frac{\partial{\bf B}}{\partial t} \, ,
\label{EQw2.08}
\end{equation}
Faraday's law, and Gauss's law into Eq.~(\ref{EQw2.07}) yields the
continuity equation
$$
\rho_f{\bf E}+\frac{{\bf J}_f}{c}\times{\bf B} =
(\nabla\cdot{\bf D}){\bf E} +
$$
\begin{equation}
(\nabla\cdot{\bf B}){\bf H} 
-{\bf D}\times(\nabla\times {\bf E})
-{\bf B}\times(\nabla\times{\bf H})
-\frac{1}{c}\frac{\partial}{\partial t}({\bf D}\times{\bf B}) \, .
\label{EQw2.09}
\end{equation}
This result is equal to Eq.~(\ref{EQw2.04}) and is also a valid theorem
of Maxwell--Minkowski electrodynamics.
Dropping the free charges and free charge currents, one reproduces the
momentum continuity equation for a neutral medium, Eq.~(\ref{EQw2.05b}).
\par
Textbook derivations \cite{BIMar,BIGriff,BIJackson,BIZangwill} of the
electromagnetic momentum continuity equation typically begin with 
the Lorentz force law, as shown by the derivation,
Eq.~(\ref{EQw2.06})--(\ref{EQw2.09}).
The derivation is simple and obvious.
However, one starts with ${\bf f}_L=0$ in the absence of free charges and
free charge currents.
This might lead one to think that free charges and free charge currents are
necessary to the theory.
This opinion is clearly wrong because free charges and free charge currents
are initial conditions and having no free charges and no free charge
current does not make the homogeneous theory wrong.
The momentum continuity equation can be derived for a neutral dielectric
medium using either of the two procedures by setting $\rho_f=0$ and
${\bf J}_f=0$ at any point in the derivations.
We can also substitute homogeneous Maxwell--Minkowski equations into
the calculus identity, Eq.~(\ref{EQw2.08}), and derive an identical result.
Whatever course we chart, we can study the physics of a free charge-free
and free charge current-free medium and add the free charges and free
charge currents to the theory when they are part of the specific system
being studied.
We also note that the postulated Lorentz force density,
Eq.~(\ref{EQw2.06}), is derived in Eq.~(\ref{EQw2.04}).
At the moment, we should not get distracted by peripheral issues.
\par
A quasimonochromatic field is an arbitrarily long, finite, constant
amplitude, unchirped pulse (square, rectangular, or top-hat pulse)
with a short (relative to the pulse length) smooth turn-on transition and
a short smooth turn-off transition.
In order to be concise and avoid an unnecessarily complicated
presentation, we adopt the plane wave-limit in which the amplitude of
the field is spatially constant over an arbitrarily large cross-sectional
area of the propagating field and then smoothly decreases at least
quadratically in the transverse spatial distance.
The phase front is constant across the transverse cross-section of the
propagating field.
The plane-wave limit is a useful concept that allows us to treat the
dynamics by a one-dimensional model as long as the well-known
characteristics are applied consistent with the well-known limits,
see, for instance, Sec.~7.1 of Ref.~\cite{BIJackson} or Chap. 16
of Ref.~\cite{BIZangwill}.
It should be noted that the plane-wave limit is distinct from the
assumption of infinite plane waves that have infinite energy.
\par
The permittivity and permeability of a linear medium are functions of the
frequency of light as indicated by the Kramers--Kronig relations.
If the center frequency $\omega_p$ of the quasimonochromatic field is far
from any material resonances then absorption can be treated as negligible
and dispersion can be treated in the lowest order of approximation by 
using the permittivity $\varepsilon(\omega_p)$ and permeability
$\mu(\omega_p)$ that correspond to the center frequency of the
quasimonochromatic field.
Many authors treat dispersion in lowest order in addressing electromagnetic
momentum issues \cite{BIObuk2} while other authors retain additional
orders \cite{BIMuka2,BIAMC2,BIAMC3,BIAMC4}.
We consider the higher orders of dispersion to be negligible or
perturbative in the parameter regime under consideration.
If it is indeed necessary to retain higher orders of dispersion, we will
still need the lowest-order theory in order to identify the source and
magnitude of any differences.
Consequently, it is not an error to correctly derive the lowest-order
theory.
\par
We consider a quasimonochromatic field with center frequency $\omega_p$
to be normally incident from the vacuum onto a finite block of a
transparent isotropic homogeneous linear medium.
The permittivity and permeability of the simple linear medium are
characterized by real, time-independent, single-valued constants
$\varepsilon(\omega_p)$ and $\mu(\omega_p)$.
Without loss of generality, we can treat the medium as being
initially at rest with respect to the Laboratory Frame of Reference.
With the medium at rest in the local frame of reference,
the macroscopic electric and magnetic fields are related by the
constitutive relations
\begin{subequations}
\begin{equation}
{\bf D}= \varepsilon {\bf E}
\label{EQw2.10a}
\end{equation}
\begin{equation}
{\bf B}= \mu {\bf H} \, ,
\label{EQw2.10b}
\end{equation}
\label{EQw2.10}
\end{subequations}
where $\varepsilon({\bf r},t_0)$ is the electric permittivity and
$\mu({\bf r},t_0)$ is the magnetic permeability.
Generally, we will treat the stationary block as a right rectangular block
of finite size that is draped with a thin gradient-index antireflection
coating, but is otherwise isotropic and homogenous.
The spatial variation of the material parameters,
$\varepsilon({\bf r},t_0)$
and $\mu({\bf r},t_0)$, typically consists of step functions (piecewise
homogeneous block material) or Fermi distributions (piecewise homogeneous
block material draped with a thin gradient-index antireflection coating).
We adopt the plane-wave limit.
Fig.~1 is a one-dimensional representation of the initial
configuration of a quasimonochromatic field that is propagating toward 
a neutral (no free charges or free charge currents), gradient-index
antireflection coated, arbitrarily long, stationary block
of transparent, homogeneous, isotropic, linear material with refractive
index $n({\bf r},t_0)=\sqrt{\varepsilon({\bf r},t_0) \mu({\bf r},t_0)}$.
\par
As the field enters the medium from the vacuum, the field imparts optically
induced surface and volume forces to the material that act to accelerate
the material, and/or portions of the material.
The nature of the surface and volume forces (Fresnel, Lorentz, Helmholtz,
Abraham, etc.), has been debated in the scientific literature for a very
long time.
However, it is the consequence of material motion on the optical
characteristics of simple linear media that is important to discuss here.
\par
Laue \cite{BILaue,BILaue2} applied the Einstein relativistic velocity sum
rule to derive the speed of light in a transparent block of dielectric that
is moving in the Laboratory Frame of Reference with velocity ${\textbf v}$.
Laue's formula for the speed of light in the moving dielectric medium is
\begin{equation}
w^{\prime}
=\frac{ \sqrt{v^2+\frac{c^2}{n^2}+2v\frac{c}{n}\cos\theta
-\frac{v^2}{n^2}\sin^2\theta}
}{
1+\frac{v}{cn}\cos\theta
} \, ,
\label{EQw2.11}
\end{equation}
where $n=\sqrt{\varepsilon}$ is the index of refraction in the rest frame
and $\theta$ is the angle between the direction of light propagation and
the direction in which the dielectric is moving \cite{BILaue}.
Then 
\begin{equation}
n^{\prime} =\frac{c}{w^{\prime}}
\label{EQw2.12}
\end{equation}
is the index of refraction in the moving frame.
However, it takes an intense light field applied for a long time for a
macroscopic material to be accelerated to relativistic speeds
where the difference between $n$ and $n^{\prime}$ would be appreciable.
Physically, Eqs.~(\ref{EQw2.10}) are valid limiting cases and are usually
very accurate.
Ramos, Rubilar, and Obukhov \cite{BIObuk} use conservation of the
center of energy velocity and also conclude that
\begin{equation}
w^{\prime} \approx w=\frac{c}{n}
\label{EQw2.13}
\end{equation}
``is an extremely accurate approximation indeed''.
\par
Describing the theoretical viewpoint of physics, Rindler \cite{BIRindler}
states ``a physical theory is an abstract mathematical model (much like
Euclidian geometry) whose applications to the real world consist of
correspondences between a subset of it and a subset of the real
world''.
Experimentalists, developers, and other realists may disagree and want to
include all potentially relevant aspects of the physical world.
However, adding complexity introduces additional parameters that are
not independently determinable making it impossible to prove or disprove a
particular model, e.g., the Abraham momentum or the Minkowski momentum.
\par
By choosing to work in a regime in which higher-than-first-order-dispersion
is negligible and the motion of the medium is non-relativistic and also
negligible, we have a rock-solid basis for our theory in terms of the
constitutive relations, ${\bf D}=\varepsilon{\bf E}$ and
${\bf B}=\mu{\bf H}$.
\par
At optical frequencies, the magnetic permeability is usually negligible and
the large majority of work on the Abraham--Minkowski dilemma has been
performed for dielectric media.
In order to maintain contact with the prior work, we restrict ourselves to
dielectric media and designate
\begin{subequations}
\begin{equation}
{\bf D}= n^2{\bf E}
\label{EQw2.14a}
\end{equation}
\begin{equation}
{\bf B}= {\bf H} 
\label{EQw2.14b}
\end{equation}
\label{EQw2.14}
\end{subequations}
as axioms of our formal theory.
These axioms are the same as the constitutive relations,
Eqs.~(\ref{EQw2.10}), but for the simpler case of a dielectric medium, with
$n=\sqrt{\varepsilon}$ and $\mu=1$, rather than the more general
magneto-dielectric medium.
\par
Using the axioms, Eqs.~(\ref{EQw2.14}), to eliminate ${\bf D}$ in favor of
${\bf E}$ and ${\bf H}$ in favor of ${\bf B}$, the energy continuity
equation, Eq.~(\ref{EQw2.05a}), can be written as
\begin{equation}
\frac{\partial}{\partial t}
\left (\frac{1}{2}\left (n^2{\bf E}^2
+{\bf B}^2 \right )\right ) 
+c\nabla\cdot ( {\bf E}\times{\bf B})= 0 \, .
\label{EQw2.15}
\end{equation}
This equation has been derived as a mathematical identity of the
macroscopic Maxwell equations for the specific case of a 
simple linear dielectric medium.
We denote the macroscopic electromagnetic energy density
\begin{equation}
\rho_e=(1/2)\left ( n^2{\bf E}^2+{\bf B}^2 \right )
\label{EQw2.16}
\end{equation}
and the Poynting energy flux vector
\begin{equation}
{\bf S}=c{\bf E}\times{\bf B} \, .
\label{EQw2.17}
\end{equation}
Substituting the two quantities, Eq.~(\ref{EQw2.16}) and (\ref{EQw2.17}),
into Eq.~(\ref{EQw2.15}), one obtains Poynting’s theorem
\begin{equation}
\frac{\partial \rho_e}{\partial t}+\nabla\cdot {\bf S}=0 \, .
\label{EQw2.18}
\end{equation}
Poynting's theorem, Eq.~(\ref{EQw2.18}), is a valid theorem of the
formal theory of Maxwellian continuum electrodynamics for a stationary
simple linear dielectric medium, as is Eq.~(\ref{EQw2.15}).
For a dielectric moving at velocity ${\bf v}$ in a Laboratory Frame of
Reference, the relativistic corrections are of order $|{\bf v}|/c$.
\par
Similarly, one can substitute the constitutive relation axioms,
Eqs.~(\ref{EQw2.14}), into the momentum continuity equation,
Eq.~(\ref{EQw2.05b}), and produce
$$
\frac{1}{c}\frac{\partial}{\partial t}(n^2{\bf E}\times{\bf B})
+n^2{\bf E}\times(\nabla\times {\bf E})
+{\bf B}\times(\nabla\times{\bf B})
$$
\begin{equation}
-n^2{\bf E}(\nabla\cdot{\bf E})
 - (\nabla\cdot{\bf B}){\bf B} = 
{\bf E}({\bf E}\cdot\nabla(n^2))
 \, .
\label{EQw2.19}
\end{equation}
The identity
$$
\left [{\bf E}(\nabla\cdot{\bf E})
-{\bf E}\times(\bf \nabla\times{\bf E})\right ]_{i}=
$$
\begin{equation}
\sum_{j}\frac{\partial}{\partial x_{j}}
\left (E_{i} E_{j} -\frac{1}{2} {\bf E}\cdot {\bf E}
\delta_{ij} \right )
\label{EQw2.20}
\end{equation}
is derived by expanding the vector operators,
See Sec. 6.8 of Ref.~\cite{BIJackson}.
Deriving a similar equation for the ${\bf B}$ field, multiplying
Eq.~(\ref{EQw2.20}) by $n^2$, and substituting the intermediate results
into the momentum continuity equation, Eq.~(\ref{EQw2.19}) produces
$$
\frac{1}{c}\frac{\partial}{\partial t}(n^2{\bf E}\times{\bf B})_{i}+
n^2\sum_{j}\frac{\partial}{\partial x_{j}}
\left (E_{i} E_{j} -\frac{1}{2} {\bf E}\cdot {\bf E}
\delta_{ij} \right )
$$
\begin{equation}
+\sum_{j}\frac{\partial}{\partial x_{j}}
\left (B_{i} B_{j} -\frac{1}{2} {\bf B}\cdot{\bf B}
\delta_{ij} \right )
={\bf E}_{i}({\bf E}\cdot\nabla(n^2))
\label{EQw2.21}
\end{equation}
for a simple linear dielectric.
Then
$$
\frac{1}{c}\frac{\partial}{\partial t}(n^2{\bf E}\times{\bf B})_{i}+
\sum_{j}\frac{\partial}{\partial x_{j}}
\left (n^2E_{i} E_{j} -\frac{1}{2}n^2 {\bf E}\cdot{\bf E}
\delta_{ij} \right )
$$
$$
+\sum_{j}\frac{\partial}{\partial x_{j}}
\left (B_{i} B_{j} -\frac{1}{2} {\bf B}\cdot{\bf B}
\delta_{ij} \right )
$$
\begin{equation}
={\bf E}({\bf E}\cdot\nabla(n^2))-
\sum_{j}\left ( \frac{\partial (n^2)}{\partial x_{j}} \right ) 
\left (n^2E_{i} E_{j} -\frac{1}{2}n^2 {\bf E}\cdot{\bf E}
\delta_{ij} \right ) 
\label{EQw2.22}
\end{equation}
by commutation.
Denoting
\begin{equation}
{{\sf W}_M}_{ij}=-n^2E_i E_j-B_iB_j+
\frac{1}{2} (n^2{\bf E}\cdot{\bf E}+{\bf B}\cdot{\bf B})\delta_{ij}
\label{EQw2.23}
\end{equation}
allows construction of another valid theorem of Maxwellian continuum
electrodynamics \cite{BIJackson}
$$
\frac{1}{c}\frac{\partial}{\partial t}(n^2{\bf E}\times{\bf B})_{i}
+\sum_{j}\frac{\partial}{\partial x_{j}}
{{\sf W}_M}_{ij} =
$$
\begin{equation}
={\bf E}({\bf E}\cdot\nabla(n^2))-
\sum_{j}\frac{\partial (n^2)}{\partial x_{j}}
 \left (n^2E_{i}E_{j}
+\frac{n^2}{2}({\bf E}\cdot{\bf E})\delta_{ij}\right ) \, ,
\label{EQw2.24}
\end{equation}
for a non-relativistic simple linear dielectric medium.
The Minkowski force density
$$
{{\bf f}_M}={\bf E}({\bf E}\cdot\nabla(n^2))-
$$
\begin{equation}
\sum_{j}\frac{\partial (n^2)}{\partial x_{j}}
\left (n^2E_{i}E_{j}+\frac{1}{2}(n^2{\bf E}
\cdot{\bf E})\delta_{ij}
\right )
\label{EQw2.25}
\end{equation}
is derived directly from the Maxwell--Minkowski equations for a
dielectric.
The Minkowski force density reduces to 
\begin{equation}
{{\bf f}_M}=-\frac{n^2}{2}\frac{\partial n^2}{\partial x_3}
|{\bf E}|^2
\label{EQw2.26}
\end{equation}
in the plane-wave limit.
\par
As a matter of linear algebra, we can write, row-wise, the energy
continuity equation, Eq.~(\ref{EQw2.15}), and the three scalar differential
equations that comprise the vector momentum continuity equation,
Eq.~(\ref{EQw2.24}), as a differential equation
\begin{equation}
\partial_{\beta}{\sf T}_M^{\alpha\beta}=f^{\alpha}_M \,,
\label{EQw2.27}
\end{equation}
where $\partial_{\beta}$ is the usual four-divergence operator
defined by 
\begin{equation}
\partial_{\beta}=
\left ( \frac{1}{c}\frac{\partial}{\partial t},
\frac{\partial}{\partial x},\frac{\partial}{\partial y},
\frac{\partial}{\partial z} \right ) ,
\label{EQw2.28}
\end{equation}
$$
f^{\alpha}_M= (0,{\bf f}_M)=\Bigg ( 0, 
{\bf E}({\bf E}\cdot\nabla(n^2))-
$$
\begin{equation}
\sum_j\frac{\partial(n^2)}{\partial x_j}
\left (n^2E_iE_j+\frac{1}{2}(n^2{\bf E}
\cdot{\bf E})\delta_{ij}\right )
\label{EQw2.29}
\end{equation}
is the Minkowski four-force density, and
\begin{equation}
{\sf T}_M^{\alpha\beta}
\! = \!
\left [
\begin{matrix}
\frac{1}{2}(n^2{\bf E}^2+{\bf B}^2)
\!&({\bf E}\times{\bf B})_1 \!& ({\bf E}\times{\bf B})_2 
\!&({\bf E}\times{\bf B})_3
\cr
(n^2{\bf E}\times{\bf B})_1    &{{\sf W}_M}_{11}      &{{\sf W}_M}_{12}      &{{\sf W}_M}_{13}
\cr
(n^2{\bf E}\times{\bf B})_2    &{{\sf W}_M}_{21}      &{{\sf W}_M}_{22}      &{{\sf W}_M}_{23}
\cr
(n^2{\bf E}\times{\bf B})_3    &{{\sf W}_M}_{31}      &{{\sf W}_M}_{32}      &{{\sf W}_M}_{33}
\cr
\end{matrix}
\right ] 
\label{EQw2.30}
\end{equation}
is, by construction, a four-by-four matrix.
The differential equation, Eq.~(\ref{EQw2.27}), is a valid theorem of the
formal theory of Maxwellian continuum electrodynamics for a simple linear
dielectric medium in the nonrelativistic limit. 
Obviously, the intent is to identify the four-by-four matrix,
Eq.~(\ref{EQw2.30}), with the Minkowski energy--momentum tensor.
\par
Next, we use the formal theory of Maxwellian continuum electrodynamics to
rigorously derive the Abraham energy--momentum theory \cite{BIAbr} that was
contemporaneous with the Minkowski theory \cite{BIMin}.
We subtract a force density-like term
\begin{equation}
{\bf f}_{A}=
\frac{\partial}{\partial t}
\frac{(n^2-1)({\bf E}\times{\bf B})}{c}
\label{EQw2.31}
\end{equation}
from both sides of Eq.~(\ref{EQw2.24}) to obtain
$$
\frac{1}{c}
\frac{\partial}{\partial t}
({\bf E}\times{\bf B})_{i}
+\sum_{j}\frac{\partial}
{\partial x_{j}}{{\sf W}_M}_{{i}{j}} =
{\bf E}_i({\bf E}\cdot\nabla(n^2))-
$$
\begin{equation}
\sum_{j}\frac{\partial (n^2)}{\partial x_{j}}
 \left (n^2E_{i}E_{j}
+\frac{n^2}{2}({\bf E}\cdot{\bf E})\delta_{{i}{j}}\right ) 
-\frac{\partial}{\partial t}
\frac{(n^2-1)({\bf E}\times{\bf B})_i}{c}\, .
\label{EQw2.32}
\end{equation}
We combine, row-wise, the energy continuity equation, Eq.~(\ref{EQw2.15}),
with the three orthogonal components of the momentum continuity equation,
Eq.~(\ref{EQw2.32}), to obtain a new differential equation
\begin{equation}
\partial_{\beta}{\sf T}_A^{\alpha\beta}=f^{\alpha}_A 
\label{EQw2.33}
\end{equation}
that is also a valid theorem of the formal theory of Maxwellian
continuum electrodynamics, where
\begin{equation}
{\sf T}_A^{\alpha\beta}
\! = \!
\left [
\begin{matrix}
\frac{1}{2}(n^2{\bf E}^2+{\bf B}^2)
\!&({\bf E}\times{\bf B})_1 \!& ({\bf E}\times{\bf B})_2 
\!&({\bf E}\times{\bf B})_3
\cr
({\bf E}\times{\bf B})_1    &{{\sf W}_M}_{11}      &{{\sf W}_M}_{12}      &{{\sf W}_M}_{13}
\cr
({\bf E}\times{\bf B})_2    &{{\sf W}_M}_{21}      &{{\sf W}_M}_{22}      &{{\sf W}_M}_{23}
\cr
({\bf E}\times{\bf B})_3    &{{\sf W}_M}_{31}      &{{\sf W}_M}_{32}      &{{\sf W}_M}_{33}
\cr
\end{matrix}
\right ] 
\label{EQw2.34}
\end{equation}
is a traceless diagonally symmetric four-by-four matrix and
\begin{equation}
f_A^{\alpha}
=\left (0,f_M^{\alpha} - \frac{\partial}{\partial t} 
\frac{(n^2-1){\bf E}\times{\bf B}}{c} \right )
\label{EQw2.35}
\end{equation}
is the Abraham four-force density.
For historical reasons, the four-by-four matrix, Eq.~(\ref{EQw2.34}), is
known as the Abraham energy--momentum tensor.
\par
It is often claimed in the scientific literature that the Abraham
four-force density, Eq.~(\ref{EQw2.35}), is negligible or ``almost''
negligible because the time average of ${\bf f}_{A}$,
Eq.~(\ref{EQw2.31}), is essentially zero due to the
oscillating field \cite{BIBrev2,BIBrevnew,BITime1,BITime2,BITime3}.
See also page 205 of Ref.~\cite{BIMol}.
However, the force density-like-term, Eq.~(\ref{EQw2.31}), cannot
``fluctuate out'' because that would mean that the first term in
Eq.~(\ref{EQw2.32}) is also negligible --- obviously, this term is
necessary for electromagnetic fields to propagate through the medium.
\par
In this section, the usual, well-known momentum continuity equations,
Eqs.~(\ref{EQw2.27}) and (\ref{EQw2.33}), have been derived from the
macroscopic Maxwell--Minkowski equations.
The axioms and conditions have been explicitly stated and the steps of
the derivation have been kept small and explicitly documented
in order to forestall any arguments about the validity of the derivation
and results.
\par
\section{Spacetime Conservation Laws}
\par
The fundamental physical principles of conservation of mass, conservation
of linear momentum, conservation of angular momentum, and conservation of
total (kinetic+potential) energy were well-established long before Maxwell
and Laue.
In continuum dynamics (fluid dynamics, for example) a continuity equation
reflects the conservation of a scalar property of an unimpeded (no external
forces, pressures, or constraints), inviscid, incoherent flow of
non-interacting particles (dust, fluid, etc.) in the continuum limit in
terms of the equality of the net rate of flux out of the otherwise empty
volume and the time rate of change of the property density
field \cite{BIFox}.
For a conserved scalar property, the continuity equation of the generic
property density $\rho$ with velocity field ${\bf u}$,
\begin{equation}
\frac{\partial\rho}{\partial t}+\nabla\cdot (\rho{\bf u})=0 \, ,
\label{EQw3.01}
\end{equation}
is derived by applying the divergence theorem to a Taylor series expansion
of the property density field $\rho$ and the scalar components of the
property flux density field
\begin{equation}
{\bf g}=\rho{\bf u}
\label{EQw3.02}
\end{equation}
to unimpeded non-relativistic flow of non-interacting particles in an
otherwise empty volume \cite{BIFox}.
\par
For unimpeded flow of mass-bearing particles in a thermodynamically closed
system, we have a conserved scalar property, the total mass,
$\int_{\Sigma} \rho_m \, dv$ that is obtained by integrating the mass 
density $\rho_m$ over the total volume $\Sigma$.
The corresponding continuity equation is
\begin{equation}
\frac{\partial\rho_m}{\partial t}+\nabla\cdot (\rho_m{\bf u})=0 \, .
\label{EQw3.03}
\end{equation}
We have a conserved vector quantity, the total momentum,
$\int_{\Sigma} \rho_m{\bf u}\, dv$, belonging to the same 
thermodynamically closed system.
The vector momentum continuity equation can be written in component form as
\begin{equation}
\frac{\partial\rho_m{\bf u}_i}{\partial t}
+{\bf u}_i(\nabla\cdot(\rho_m {\bf u}))=0 \, .
\label{EQw3.04}
\end{equation}
As a matter of linear algebra, we can write, row-wise,
Eq.~(\ref{EQw3.03}) and the three scalar differential equations that
comprise the vector differential equation, Eq.~(\ref{EQw3.04}), as a
single differential equation
\begin{equation}
\partial_{\beta}{\sf T}_m^{\alpha\beta}=0 \, ,
\label{EQw3.05}
\end{equation}
where
\begin{equation}
{\sf T}_m^{\alpha\beta}
\! = \!
\left [
\begin{matrix}
\rho_m c^2       &\rho_m c u_1       &\rho_m c u_2        &\rho_m c u_3 
\cr
\rho_m c u_1     &\rho_m  u_1 u_1   &\rho_m  u_1 u_2    &\rho_m  u_1 u_3
\cr
\rho_m c u_2     &\rho_m  u_2 u_1   &\rho_m  u_2 u_2    &\rho_m  u_2 u_3
\cr
\rho_m c u_3     &\rho_m  u_3 u_1   &\rho_m  u_3 u_2    &\rho_m  u_3 u_3
\cr
\end{matrix}
\right ] 
\label{EQw3.06}
\end{equation}
is, by construction, a diagonally symmetric four-by-four matrix.
The differential equation, Eq.~(\ref{EQw3.05}), is a valid theorem of the
formal theory of continuum dynamics (not continuum electrodynamics).
Obviously, the intent is to identify the matrix, Eq.~(\ref{EQw3.06}), with
the dust energy--momentum tensor.
\par
The matrix, Eq.~(\ref{EQw3.06}), has the following characteristics of flow
through an otherwise empty volume for an unimpeded (no external forces,
pressures, or constraints), inviscid, incoherent flow of non-interacting
particles in the continuum limit \cite{BIPfei,BIFox,BILL} 
\par
1) Continuity equations (local conservation laws) are generated by the
four-divergence of the matrix \cite{BIFox,BIGiu},
\begin{equation}
\partial_{\beta}{\sf T}_m^{\alpha\beta}=0\, ,
\label{EQw3.07}
\end{equation}
by construction.
The mass continuity equation, Eq.~({\ref{EQw3.03}), is associated
with the first row of the matrix in Eq.~(\ref{EQw3.06}) via
$\partial_{\beta} {\sf T}_m^{0\beta}=0$.
Similarly, the components of the momentum continuity equation
Eq.~(\ref{EQw3.04}) are related to the other three rows of the matrix
because $\partial_{\beta} {\sf T}_m^{i\beta}=0$.
\par
2) For unimpeded, inviscid, incoherent fluid flow in the absence of sources
or sinks, the mass and momentum are globally conserved.
Then \cite{BILL,BIGiu}
\begin{equation}
P^{\alpha}(t)=
\int_{\Sigma}{\sf T}_m^{\alpha 0}dv=
P^{\alpha}(t_0)
\label{EQw3.08}
\end{equation}
is constant in time for each $\alpha$.
\par
3) The trace ${\sf T}^{\alpha \alpha}$ is proportional to the mass density
$\rho_m$:
\begin{equation}
{\sf T}^{\alpha \alpha}
\propto \rho_m \, .
\label{EQw3.09}
\end{equation}
\par
4) The matrix is diagonally symmetric,
\begin{equation}
{\sf T}_m^{\alpha\beta}= {\sf T}_m^{\beta\alpha} \, . 
\label{EQw3.10}
\end{equation}
Symmetry putatively corresponds to conservation of angular
momentum \cite{BIObuk2},
although symmetry is not considered to be an absolute requirement for
angular momentum conservation \cite{BILL}.
\par
5) The condition
\begin{equation}
\partial_{\alpha}{\sf T}_m^{\alpha\beta}=0
\label{EQw3.11}
\end{equation}
can be derived from Eq.~(\ref{EQw3.07}) using the symmetry
condition, Eq.~(\ref{EQw3.10}).
\par
With the advent of relativity, conservation of mass became conservation of
relativistic mass-energy $E=({\bf p}\cdot{\bf p}c^2+m^2c^4)^{1/2}$ and
these physical principles are known as the spacetime conservation laws and
are properties of Minkowski spacetime.
Mass-energy is simply the most well-known of the conserved properties: the
discussion in this section applies, equally well, for the conservation of
number/quantity and for the conservation of any intrinsic property of 
identical non-interacting particles in an unimpeded, inviscid, incoherent
flow through empty space.
\par
Although the material that is presented in this section is well-known,
some experts have questioned the application of the ``particle''
conservation laws to electrodynamics.
However, the fundamental basis of the conservation laws is Minkowski
spacetime, not particle dynamics.
Application of the spacetime conservation laws to the continuum limit
of the flow of photons is demonstrated in the next section.
\par
\section{Electromagnetic Conservation}
\par
\subsection{Vacuum}
\par
The energy and momentum conservation properties of a continuous light field
propagating in the vacuum were long-ago cast in the energy--momentum tensor
formalism of classical particle dynamics in the continuum limit
in which the continuous light field plays the role of the continuous
fluid \cite{BILL}.
Because the conservation properties of light in a dielectric remain
contentious, we should reproduce the vacuum theory so that we can agree
on the terminology, procedures, and principles.
\par
The Maxwell equations for electromagnetic fields in the vacuum are
\begin{subequations}
\begin{equation}
\nabla\times{\bf e}+\frac{1}{c}\frac{\partial{\bf b}}{\partial t}=0
\label{EQw4.01a}
\end{equation}
\begin{equation}
\nabla\times{\bf b}-\frac{1}{c}\frac{\partial{\bf e}}{\partial t}=0
\label{EQw4.01b}
\end{equation}
\begin{equation}
\nabla\cdot{\bf b}=0
\label{EQw4.01c}
\end{equation}
\begin{equation}
\nabla\cdot{\bf e}=0 
\label{EQw4.01d}
\end{equation}
\label{EQw4.01}
\end{subequations}
in terms of the microscopic electric and magnetic fields,
${\bf e}({\bf r},t)$ and ${\bf b}({\bf r},t)$.
The microscopic Maxwell equations, Eqs.~(\ref{EQw4.01}) can be
systematically combined (like in Sec.~2) to form a scalar energy
continuity equation
\begin{equation}
\frac{1}{c}\left (
{\bf e}\cdot\frac{\partial{\bf e}}{\partial t}
+{\bf b}\cdot \frac{\partial{\bf b}}{\partial t}
\right )
+\nabla\cdot ( {\bf e}\times{\bf b})=0
\label{EQw4.02}
\end{equation}
and the components of a vector momentum continuity
equation \cite{BIJackson}
\begin{equation}
\frac{1}{c}\frac{\partial}{\partial t}({\bf e}\times{\bf b})_i
+\sum_j\frac{\partial}{\partial x_j}{{\sf W}_v}_{ij}
=0 \, ,
\label{EQw4.03}
\end{equation}
where
\begin{equation}
{{\sf W}_v}_{ij}=-e_i e_j-b_ib_j+
\frac{1}{2} ({\bf e}\cdot{\bf e}+{\bf b}\cdot{\bf b})\delta_{ij} \, .
\label{EQw4.04}
\end{equation}
These energy and momentum time-evolution equations can be
combined, row-wise, to construct a differential equation
\begin{equation}
\partial_{\beta}{\sf T}_v^{\alpha\beta}=0 \, ,
\label{EQw4.05}
\end{equation}
where
\begin{equation}
{\sf T}_v^{\alpha\beta}
\! = \!
\left [
\begin{matrix}
\frac{1}{2}({\bf e}^2+{\bf b}^2)
\!&({\bf e}\times{\bf b})_1 \!& ({\bf e}\times{\bf b})_2
\!&({\bf e}\times{\bf b})_3
\cr
({\bf e}\times{\bf b})_1    &{{\sf W}_v}_{11}      &{{\sf W}_v}_{12}      &{{\sf W}_v}_{13}
\cr
({\bf e}\times{\bf b})_2    &{{\sf W}_v}_{21}      &{{\sf W}_v}_{22}      &{{\sf W}_v}_{23}
\cr
({\bf e}\times{\bf b})_3    &{{\sf W}_v}_{31}      &{{\sf W}_v}_{32}      &{{\sf W}_v}_{33}
\cr
\end{matrix}
\right ] 
\label{EQw4.06}
\end{equation}
is the energy--momentum tensor for the electromagnetic field in free space.
\par
1) By construction, the continuity equation, Eq.~(\ref{EQw4.05}), is a
valid theorem of Eqs.~(\ref{EQw4.01}) that expresses \textit{local}
conservation of energy and momentum \cite{BIGiu}.
\par
2) The Laue theorem \cite{BIGiu} defines the conditions under which a
local distribution of energy and momentum can be used to construct globally
conserved quantities. 
We take the temporal constancy of 
\begin{equation}
P^{\alpha}(t)= \int_{\Sigma}{\sf T}_v^{\alpha 0}dv= P^{\alpha}(t_0)
\label{EQw4.07}
\end{equation}
for each $\alpha$ as an operational condition for global conservation of
energy and momentum for quasimonochromatic fields.
\par
3) The matrix, Eq.~(\ref{EQw4.06}), is traceless
\begin{equation}
{\sf T}_v^{\alpha \alpha}= 0
\label{EQw4.08}
\end{equation}
corresponding to massless photons.
\par
4) The matrix is diagonally symmetric
\begin{equation}
{\sf T}_v^{\alpha \beta}= {\sf T}_v^{\beta \alpha}
\label{EQw4.09}
\end{equation}
corresponding to conservation of angular momentum, although symmetry is not
considered to be an expressly rigid requirement for angular momentum
conservation \cite{BILL}.
\par
5) The condition
\begin{equation}
\partial_{\alpha}{\sf T}_v^{\alpha\beta}=0
\label{EQw4.10}
\end{equation}
is derived from Eq.~(\ref{EQw4.05}) if the matrix, Eq.~(\ref{EQw4.06})
is symmetric.
\par
The amplitude and duration of the fields are not affected by propagation
through the vacuum in the plane-wave limit insuring global conservation of
electromagnetic energy and momentum.
Clearly, the particle description of energy--momentum conservation can be
applied to the light field in the plane-wave limit as the unimpeded,
inviscid, incoherent flow of massless photons in the continuum limit
through an otherwise empty volume.
\par
\subsection{Dielectric}
\par
Microscopically, a dielectric consists of tiny polarizable particles and
host material embedded in the vacuum.
In continuum electrodynamics, the properties of the medium are averaged and
the material is continuous at all length scales.
This is a second and distinct meaning of the word ``continuum'' in
continuum electrodynamics because the light field is the continuum limit of
the flow of photons in the sense of fluid mechanics or continuum dynamics.
\par
The material is modeled as an arbitrarily large continuous isotropic
homogeneous block of transparent linear dielectric that is draped
with a gradient-index antireflection coating.
In the limit that the gradient of the refractive index can be
neglected, the Minkowski continuity equation, Eq.~(\ref{EQw2.27})
becomes
\begin{equation}
\partial_{\beta}{\sf T}_M^{\alpha\beta}=0 \, ,
\label{EQw4.11}
\end{equation}
which is the putative condition for local conservation of energy and
linear momentum \cite{BIGiu}.
Consequently, it is frequently claimed in the scientific literature that
the Minkowski linear momentum and the Minkowski energy--momentum tensor
are (globally) conserved or nearly conserved \cite{BIPfei,BIObuk}.
The gradient nature of the Minkowski four-force density,
Eq.~(\ref{EQw2.29}), definitely supports that assertion.
At this point, we would like to make it emphatically clear that this
assertion is \textit{false}.
\par
Propagation of the electromagnetic field in a neutral transparent
dielectric is given by the wave equation
\begin{equation}
\nabla\times(\nabla\times {\bf A})
+ \frac{n^2}{c^2}\frac{\partial^2{\bf A}}{\partial t^2} =0 
\label{EQw4.12}
\end{equation}
in the quasistationary limit \cite{BIKemplatest},
where ${\bf A}$ is the vector potential and 
\begin{equation}
{\bf B}=\nabla\times{\bf A}
\label{EQw4.13}
\end{equation}
\begin{equation}
{\bf E}=-\frac{1}{c}\frac{\partial {\bf A}}{\partial t} \, .
\label{EQw4.14}
\end{equation}
\par
For quasimonochromatic fields, we define the slowly varying amplitude of
the electric field ${\bf E}_0 ({\bf r},t)$ and the slowly varying amplitude
of the magnetic field ${\bf B}_0 ({\bf r},t)$ by ${\bf E}={\bf E}_0
\left (\exp( -i(\omega t-{\bf k}\cdot{\bf r}))+c.c.\right )$
and 
${\bf B}= \left ({\bf B}_0
\exp( -i(\omega t-{\bf k}\cdot{\bf r}))+c.c.\right )$.
Also,
${\bf A}= \left ({\bf A}_0
\exp( -i(\omega t-{\bf k}\cdot{\bf r}))+c.c.\right )$.
\par
Figure~1 is a graphical representation of the slowly varying amplitude
of a quasimonochromatic field (plane-wave limit) in the vacuum traveling
to the right at some time $t_0$ before entering a dielectric medium.
The representation is in terms of the envelope of the vector potential,
$|{\bf A}_0(t_0,z)|$.
A finite-difference time-domain solution of the wave equation in
retarded time \cite{BIIcsevgiLamb} allows us illustrate the same field
at some time $t_1$ after it has entered a linear isotropic homogenous
dielectric through a gradient-index antireflection coating, 
Fig.~2.
\begin{figure}
\includegraphics[scale=0.45]{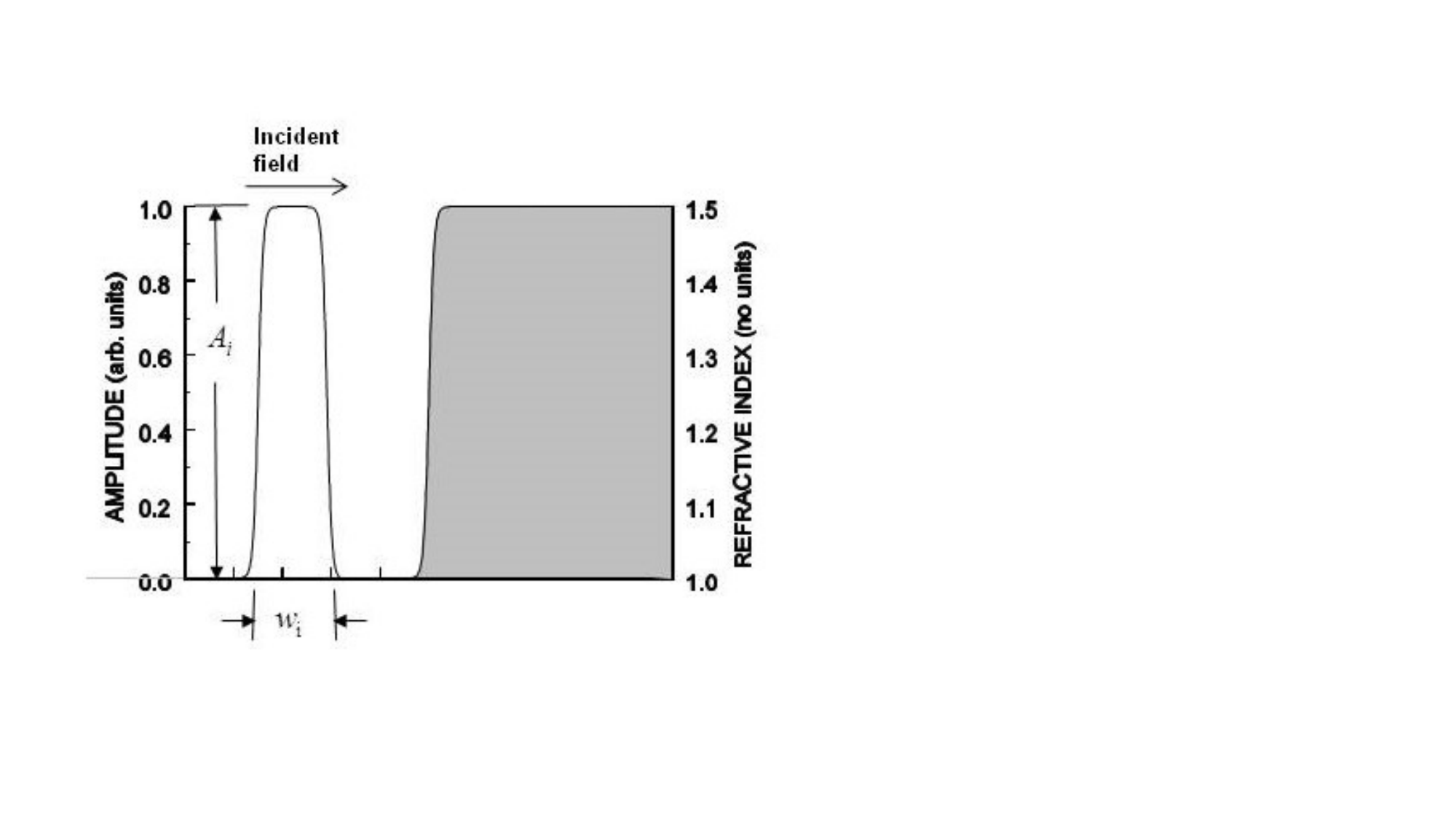}
\caption{Representation of the field in the vacuum, in terms of the
envelope of the vector potential, at some initial time $t=t_0$ before
entering the dielectric. The field, on the left, is traveling to the
right. The dielectric, draped with a gradient-index anti-reflection
coating, is represented by the outline of the refractive index on
the right.}
\label{fieldfig1}
\end{figure}
\par
According to the Maxwellian continuum electrodynamic theory, the
nominal width of the pulse in the dielectric is the width of the
incident pulse reduced by a factor of $n$ due to the reduced speed
of light in a dielectric.
The numerical solution of the wave equation confirms this theoretical
fact.
\par
The Minkowski electromagnetic energy formula is
\begin{equation}
U_M=\int_{\Sigma} {\sf T}_M^{00} \phi dz =
\int_{\Sigma} \frac{n^2{\bf E}^2+{\bf B}^2}{2} dv  \, .
\label{EQw4.15}
\end{equation}
Substituting the relations between the fields and vector potential into the
energy formula, on obtains
\begin{equation}
\int_{\Sigma} |{\bf A}_0(t_0,z)|^2 \phi dz =
\int_{\Sigma} n^2 |{\bf A}_0(t_1,z)|^2 \phi dz \, ,
\label{EQw4.16}
\end{equation}
where $\phi$ is the cross-sectional area of the field and comparisons
have a per unit of cross-sectional area basis.
This result is confirmed by numerical integration of the incident and
refracted fields shown in Figs.~1 and 2.
Using the fact that the width of the pulse is narrower in the medium by a
factor of $n$ in Eq. (\ref{EQw4.16}), one finds, by energy conservation,
that the amplitude of the vector potential in the dielectic is smaller than
the incident vector potential amplitude by a factor of $\sqrt{n}$.
This is confirmed in the numerical solution of the wave equation by 
an examination of Figs. 1 and 2.
Both theoretically and numerically it is shown that the amplitude of the
vector potential in the dielectric is the amplitude of the incident
vector potential divided by $\sqrt{n}$.
Applying this result to Eqs.~(\ref{EQw4.13}) and (\ref{EQw4.14}), one finds
that the amplitude of the electric field in the dielectric,
$|{\bf E}_0|_{diel}$, is a factor of $\sqrt{n}$ smaller than the amplitude
of the electric field that is incident from the
vacuum, $|{\bf E}_0|_{inc}$.
Meanwhile, the amplitude of the magnetic field in the dielectric is
increased by a factor of $\sqrt{n}$ from the amplitude of the incident
field.
For a quasimonochromatic field in the quasistationary limit,
\begin{subequations}
\begin{equation}
|{\bf E}_0|_{diel}= |{\bf E}_0|_{inc}/\sqrt{n}
\label{EQw4.17a}
\end{equation}
\begin{equation}
|{\bf B}_0|_{diel}= \sqrt{n}|{\bf B}_0|_{inc} \, .
\label{EQw4.17b}
\end{equation}
\label{EQw4.17}
\end{subequations}
\par
Applying the relations between the incident fields and the fields in the
dielectric, Eqs.~(\ref{EQw4.17}), we find that ${\bf E}\times{\bf B}$ has a
constant amplitude across the antireflection-coated entry face of the
dielectric.
Then, ${\bf E}\times{\bf B}$ is multiplied by $n^2/c$ to get the 
Minkowski momentum density.
The pulse is narrower in the dielectric than in the vacuum by a factor
of $n$ due to the reduced velocity of light in the dielectric.
Substituting the relations between the incident fields and the fields
in the dielectric into the Minkowski electromagnetic momentum formula
\begin{equation}
{\bf G}_M=\frac{1}{c}\int_{\Sigma}
\left ( {\sf T}_M^{10},{\sf T}_M^{20},{\sf T}_M^{30} \right ) \phi dz 
=\int_{\Sigma} \frac{n^2{\bf E}\times{\bf B}}{c} dv 
\label{EQw4.18}
\end{equation}
and comparing the results with Eqs.~(\ref{EQw4.15}) and (\ref{EQw4.16}),
we find that the Minkowski electromagnetic energy is constant in time but
the Minkowski electromagnetic momentum is greater than the incident
momentum by a factor of $n$ \textit{strongly violating} the global
conservation condition, Eq.~(\ref{EQw3.08}).
It could not be otherwise because Eqs.~(\ref{EQw4.15}) and
(\ref{EQw4.18}) have the same quadratic dependence of the fields but
differ by a factor of $n$ in magnitude.
Consequently, ${\bf G}_M(t_1)$ is very different from the incident
momentum and ${\sf T}_M^{\alpha\beta}$ is not an approximation of the total
energy--momentum tensor, contrary to the conservation implied by
Eq.~(\ref{EQw4.11}) and contrary to statements in the scientific
literature \cite{BIPfei,BIObuk}.
\begin{figure}
\includegraphics[scale=0.45]{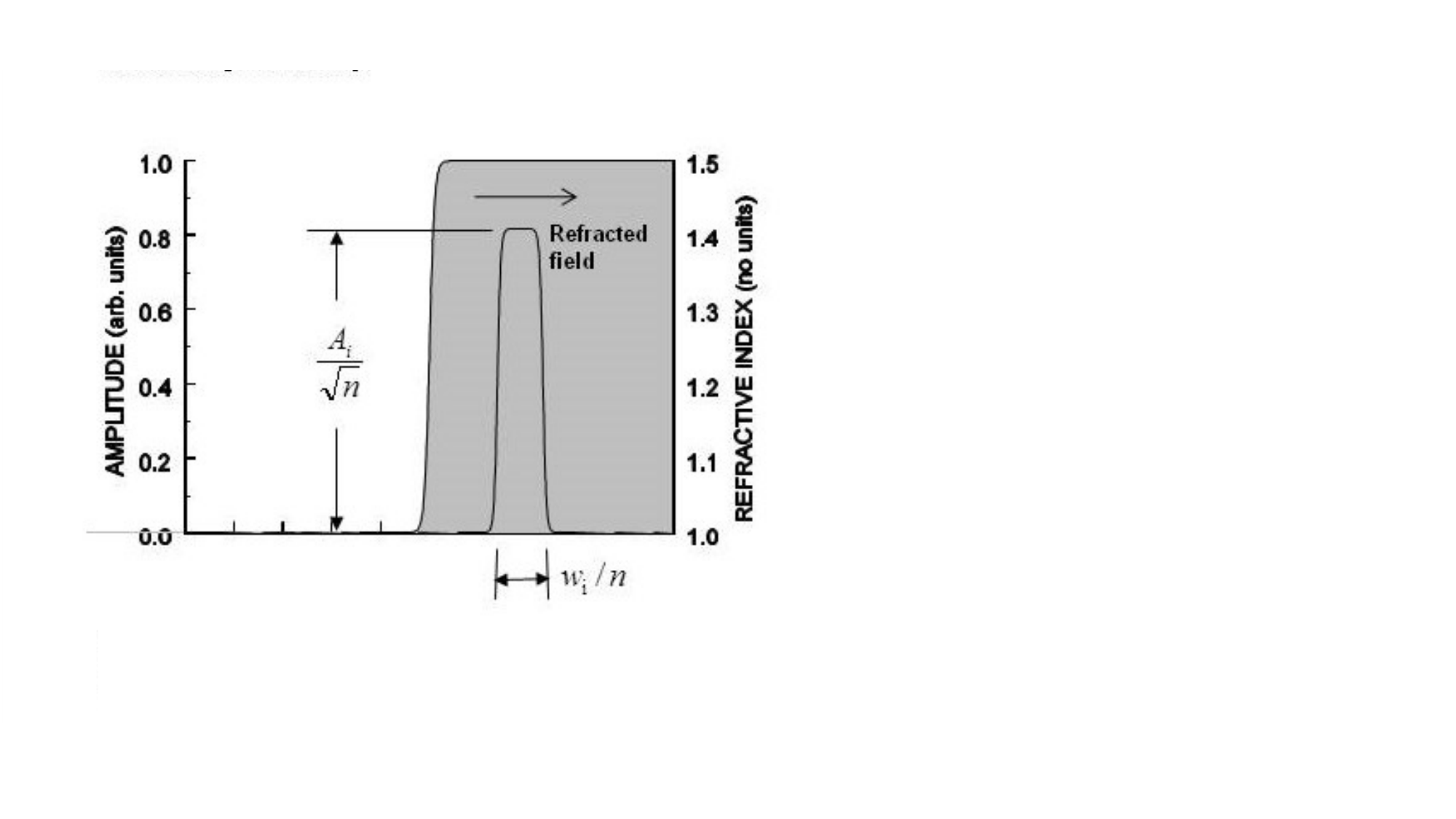}
\caption{Representation of the field, in terms of the envelope of the
vector potential, at a time $t=t_1$ when the field is entirely within
the dielectric.}
\label{fieldfig2}
\end{figure}
\par
We have shown that energy--momentum relations that are systematically
derived from the field equations using Maxwellian continuum
electrodynamics are inconsistent with spacetime conservation laws
if the gradient force is negligible as in Eq.~(\ref{EQw4.11}).
In order to avoid that fate, it has been the practice to make a
physically motivated assumption that the macroscopic Maxwell--Minkowski
equations describe an electromagnetic subsystem that is coupled to a
material subsystem.
We now examine this alternative, generally accepted as correct case 
\cite{BIRL,BIBrev,BIAMC3,BIKemplatest,BIPfei,BIAMC2,BIAMC4,BIAMC5,BIObuk,BIObuk2,BIMuka,BIMuka2,BIBarn,BIPeierls,BIBalazs,BIGord,BIMol,BIBahder,BIJMP},
and show that it also leads to strong violation of spacetime conservation
laws.
\par
For a given electromagnetic energy--momentum four-tensor
${\sf T}_{em}^{\alpha\beta}$ (Minkowski, Abraham, etc.)
there is an associated four-force density $f^{\alpha}_{em}$ such that
\begin{equation}
\partial_{\beta} {\sf T}_{em}^{\alpha\beta} = f^{\alpha}_{em} \, .
\label{EQw4.19}
\end{equation}
According to the current Abraham--Minkowski resolution theory
\cite{BIRL,BIBrev,BIAMC3,BIKemplatest,BIPfei,BIAMC2,BIAMC4,BIAMC5,BIObuk,BIObuk2,BIMuka,BIMuka2,BIBarn,BIPeierls,BIBalazs,BIGord,BIMol,BIBahder,BIJMP},
the dynamics of the material sub-system are
based on a material four-tensor ${\sf T}_{matl}^{\alpha\beta}$ such that
\begin{equation}
\partial_{\beta} {\sf T}_{matl}^{\alpha\beta} = -f^{\alpha}_{em} \, .
\label{EQw4.20}
\end{equation}
We add Eqs.~(\ref{EQw4.19}) and (\ref{EQw4.20}).
Then the total energy--momentum tensor
\begin{equation}
{\sf T}_{total}^{\alpha\beta}=
{\sf T}_{em}^{\alpha\beta}+ {\sf T}_{matl}^{\alpha\beta}
\label{EQw4.21}
\end{equation}
obeys the local conservation law \cite{BIGiu}
\begin{equation}
\partial_{\beta} {\sf T}_{total}^{\alpha\beta} =0 
\label{EQw4.22}
\end{equation}
in accordance with the spacetime conservation law Eq.~(\ref{EQw3.07}).
\par
A wide variety of physical models have been employed in an effort
to fully resolve the problem of momentum conservation in a dielectric
\cite{BIRL,BIBrev,BIAMC3,BIKemplatest,BIPfei,BIAMC2,BIAMC4,BIAMC5,BIObuk,BIObuk2,BIMuka,BIMuka2,BIBarn,BIPeierls,BIBalazs,BIGord,BIMol,BIBahder,BIJMP}.
Selected examples are discussed in the next subsection.
Typically, one assumes a microscopic model of the material dynamics in
a dielectric and applies an averaging technique to derive the
macroscopic momentum of the material.
The correctness of the results is assumed to be affirmed by the
fundamental nature of the physical laws that are used as the basis of
the analysis.
Adding the electromagnetic and material tensors, one obtains the total
energy--momentum tensor for the thermodynamically closed system, 
Eq.~(\ref{EQw4.21}).
The total linear momentum
\begin{equation}
{\bf G}_{total} =\frac{1}{c} \int_{\Sigma}
\left (
{\sf T}_{total}^{10},{\sf T}_{total}^{20},{\sf T}_{total}^{30}
\right ) \phi dz 
=\int_{\Sigma} \frac{n {\bf E}\times{\bf B} }{c} dv 
\label{EQw4.23}
\end{equation}
and the total energy
\begin{equation}
U_{total} =\int_{\Sigma} {\sf T}_{total}^{00} \phi dz
=\int_{\Sigma} \frac{n^2{\bf E}^2+{\bf B}^2}{2} dv
\label{EQw4.24}
\end{equation}
are \textit{known} quantities that can be related to the energy and
momentum of the incident field in the vacuum because they are required to
be constant in time by global conservation laws in the \textit{complete}
and \textit{closed} system (unless one assumes inappropriate system
boundaries).
Using the corresponding total energy and total momentum densities to
populate the total energy--momentum tensor, we write \cite{BIJMP}
\par
$$
{\sf T}_{total}^{\alpha\beta}
\! = \!
$$
\begin{equation}
\left [
\begin{matrix}
\frac{1}{2}(n^2{\bf E}^2+{\bf B}^2)
\!&(n{\bf E}\times{\bf B})_1 \!& (n{\bf E}\times{\bf B})_2 
\!&(n{\bf E}\times{\bf B})_3
\cr
(n{\bf E}\times{\bf B})_1    &{{\sf W}_M}_{11}      &{{\sf W}_M}_{12}      &{{\sf W}_M}_{13}
\cr
(n{\bf E}\times{\bf B})_2    &{{\sf W}_M}_{21}      &{{\sf W}_M}_{22}      &{{\sf W}_M}_{23}
\cr
(n{\bf E}\times{\bf B})_3    &{{\sf W}_M}_{31}      &{{\sf W}_M}_{32}      &{{\sf W}_M}_{33}
\cr
\end{matrix}
\right ] \, .
\label{EQw4.25}
\end{equation}
The total energy--momentum tensor, Eq.~(\ref{EQw4.25}), is diagonally
symmetric and traceless.
Both the total energy, Eq.~(\ref{EQw4.24}) and the total linear
momentum, Eq.~(\ref{EQw4.23}), are constant in time thereby satisfying
the global conservation condition, Eq.~(\ref{EQw3.08}).
Applying the local conservation condition,
Eq.~(\ref{EQw3.07}), to the total energy--momentum tensor,
Eq.~(\ref{EQw4.25}), we find that the
energy continuity equation 
\begin{equation}
\frac{1}{c}\frac{\partial}{\partial t}\frac{1}{2}
\left ( n^2{\bf E}^2+{\bf B}^2 \right ) 
+\nabla\cdot ( n{\bf E}\times{\bf B})=0 
\label{EQw4.26}
\end{equation}
that is obtained for $\alpha=0$ is
\textit{manifestly false} because the two non-zero terms depend on
different powers of $n$ in addition to being incommensurate with the
Poynting theorem.
This result is based on the total (electromagnetic plus material)
energy--momentum tensor, Eq.~(\ref{EQw4.25}), and is therefore independent
of the particular electromagnetic representation, Abraham, Minkowski, etc.,
that is used.
Then, the macroscopic Maxwell field equations and the spacetime
conservation laws are laws of physics that are proven to be
contradictory in the case of a thermodynamically closed system
consisting of an electromagnetic subsystem and a dielectric material
subsystem \cite{BIBrevnew}.
Clearly, the prescribed method to resolve the Abraham--Minkowski momentum
dilemma produces only another contradiction.
\par
It may be argued that pure induction without experimental support
is not a method of theoretical physics.
In our case, the energy--momentum evolution equations,
Eqs.~(\ref{EQw2.27}) and (\ref{EQw2.33}), are derived by formal theory
directly from the laws of Maxwell-Minkowski continuum electrodynamics.
Then the Minkowski momentum was shown to strongly violate global
conservation laws.
When these theorems are ``fixed'' by the addition of a physically
motivated, but hypothetical, material energy--momentum tensor as shown in
Eq.~(\ref{EQw4.21}), the contrived total energy--momentum tensor,
Eq.~(\ref{EQw4.25}), leads
to violation of other conditions of the spacetime conservation laws as
shown by Eq.~(\ref{EQw4.26}).
\par
Proof by mathematical contradiction is far stronger than an
experimental demonstration.
One might recall that the 1887 Michelson--Morley experiment
\cite{BIMichMor} was initially interpreted to prove the existence of ether
drag and was later deemed to support the absence of ether in the Einstein
relativity theory.
Likewise, the experiments that were originally viewed as support for the
Abraham energy--momentum theory or the Minkowski theory, or both, will
be shown in Sec.~7 to provide experimental justification for the new
theory that is derived in Sec.~5.
\par
\subsection{Brief Survey of Prior Work}
\par
The century-long history of the Abraham--Minkowski controversy
\cite{BIMin,BIAbr,BIRL,BIBrev,BIAMC3,BIKemplatest,BIPfei,BIAMC2,BIAMC4,BIAMC5,BIObuk,BIObuk2,BIMuka,BIMuka2,BIBarn,BIPeierls,BIBalazs,BIGord,BIMol,BIBahder,BIJMP}
is a search for some provable description of momentum and momentum 
conservation for electromagnetic fields in dielectric media.
A wide variety of physical principles have been applied to establish the
priority of one type of momentum over another, or to establish that the
Abraham and Minkowski formulations are equally valid.
The modern resolution of the Abraham--Minkowski momentum controversy is
to adopt a scientific conformity in which the Minkowski momentum
and the Abraham momentum are both correct forms of electromagnetic
momentum with the understanding that neither is the total
momentum \cite{BIPfei,BIAMC2,BIAMC5,BIBarn}.
Either the Minkowski momentum or the Abraham momentum can be used
as the momentum of the electromagnetic field as long as that momentum is
accompanied by the appropriate material momentum \cite{BIPfei}.
The material momentum is specific to a particular material and we will
consider several well-known models that have appeared in the scientific
literature in order to circumscribe the area of difficulty.
\par
In a quasi-microscopic approach, the material momentum is often modeled
as the aggregated kinematic momentum of individual particles of matter 
in the continuum limit.
The total energy--momentum tensor is the sum of the electromagnetic 
energy--momentum tensor and the material energy--momentum tensor.
In one example, 
Pfeifer, Nieminen, Heckenberg, and Rubinsztein-Dunlop \cite{BIPfei},
posit that the total energy--momentum tensor is the sum of the
Abraham energy--momentum tensor, Eq.~(\ref{EQw2.34}), and
the dust energy--momentum tensor
\begin{equation}
{\sf T}_{mat,Abr}^{\alpha\beta} = 
{\sf T}_{dust}^{\alpha\beta} = 
\left [
\begin{matrix}
\rho_0 c^2
\!&\rho_0 c v_1 \!&\rho_0 c v_2 
\!&\rho_0 c v_3
\cr
\rho_0 c v_1        &\rho_0 v_1 v_1  &\rho_0 v_1 v_2  &\rho_0 v_1 v_3
\cr
\rho_0 c v_2        &\rho_0 v_2 v_1  &\rho_0 v_2 v_2  &\rho_0 v_2 v_3
\cr
\rho_0 c v_3        &\rho_0 v_3 v_1  &\rho_0 v_3 v_2  &\rho_0 v_3 v_3
\cr
\end{matrix}
\right ] \, .
\label{EQw4.27}
\end{equation}
Here, $\rho_0$ is a constant mass density and ${\bf v}({\bf r},t)$ is a
velocity field.
The dust tensor, Eq.~(\ref{EQw4.27}), is usually applied to a
thermodynamically closed system consisting of non-interacting, neutral,
mass-bearing particles in an inviscid, incoherent, unimpeded flow such
that
\begin{equation}
\partial_{\beta} {\sf T}_{dust}^{\alpha\beta} = 0
\label{EQw4.28}
\end{equation}
in the continuum limit.
In the current context, however, the total tensor energy--momentum
continuity equation is posited as \cite{BIPfei}
\begin{equation}
\partial_{\beta} {\sf T}_{total}=
\partial_{\beta} \left (
{\sf T}_A^{\alpha\beta}+{\sf T}_{dust}^{\alpha\beta}
\right )=0 \,.
\label{EQw4.29}
\end{equation}
Clearly, it is intended that the dust tensor is coupled to the Abraham
electromagnetic tensor through the Abraham force density such
that \cite{BIBethune}
\begin{equation}
\partial_{\beta} {\sf T}_{dust}^{\alpha\beta} = -f^{\alpha}_A \, .
\label{EQw4.30}
\end{equation}
Performing the substitution of the dust tensor, Eq.~(\ref{EQw4.27}),
and the Abraham energy--momentum tensor, Eq.~(\ref{EQw2.34}), into
Eq.~(\ref{EQw4.29}) results in \cite{BIPfei}
\begin{equation}
\frac{1}{c}\frac{\partial}{\partial t}
\left [ \rho_0 c^2+\frac{1}{2}\left (n^2{\bf E}^2+{\bf B}^2\right )
\right ]
+\nabla\cdot(\rho_0 {\bf v})
+\nabla\cdot ( {\bf E}\times{\bf B})=0
\label{EQw4.31}
\end{equation}
for the $\alpha=0$ component of the total tensor energy--momentum
continuity equation, Eq.~(\ref{EQw4.29}).
Because this sub-section is devoted to documenting the prior work,
any issue with Eq.~(\ref{EQw4.31}) does not condemn the new work that
is presented in Secs. 5--7 of this article.
Within the context of the prior work, Eq.~(\ref{EQw4.31}) and, by
extension, Eq.~(\ref{EQw4.46}), are proven false in the next paragraph.
\par
Pfeifer, Nieminen, Heckenberg, and Rubinsztein-Dunlop \cite{BIPfei}
then use global conservation of momentum arguments to phenomenologically
relate the material momentum density to the electromagnetic momentum
density with the {\it ansatz}
\begin{equation}
\rho_0{\bf v}={\bf g}_{total}-{\bf g}_A 
=(n-1)\frac{{\bf E}\times{\bf B}}{c} \, .
\label{EQw4.32}
\end{equation} 
Substituting Eq.~(\ref{EQw4.32}) into Eq.~(\ref{EQw4.31})
produces \cite{BIPfei}
\begin{equation}
\frac{1}{c}\frac{\partial}{\partial t} \left [\rho_0 c^2
+\frac{1}{2}\left (n^2{\bf E}^2+{\bf B}^2\right )\right ]
+\nabla\cdot ( n{\bf E}\times{\bf B})=0 \, .
\label{EQw4.33}
\end{equation}
The total energy and total momentum are both quadratic in the fields
and must have the same dependence on the refractive index $n$.
We note that if the particle density $\rho_0$ is constant then
Eq.~(\ref{EQw4.33})  reproduces Eq.~(\ref{EQw4.26}) and is manifestly
false because the two nonzero terms would depend on different powers of the
refractive index and because the equation would be incommensurate with
Poynting's theorem.
Although
Pfeifer, Nieminen, Heckenberg, and Rubinsztein-Dunlop \cite{BIPfei}
do not propose a time-dependent model for the particle density $\rho_0$,
the two non-zero terms of the energy continuity equation will be
incommensurate unless Eq.~(\ref{EQw4.33}) becomes
\begin{equation}
\frac{n}{c}\frac{\partial}{\partial t}
\left [ \frac{1}{2}\left (n^2{\bf E}^2+{\bf B}^2\right )\right ]
+\nabla\cdot ( n{\bf E}\times{\bf B})=0 \, .
\label{EQw4.34}
\end{equation}
The corresponding tensor continuity equation for the total energy and
the total momentum
\begin{equation}
\bar\partial_{\beta} {\sf T}_{total}^{\alpha \beta} = 0
\label{EQw4.35}
\end{equation}
is false because the presence of the index-dependent \textit{material}
four-divergence operator \cite{BIFinn,BIJMP}
\begin{equation}
\bar\partial_{\beta} =
\left ( \frac{n}{c}\frac{\partial}{\partial t},
\frac{\partial}{\partial x},\frac{\partial}{\partial y},
\frac{\partial}{\partial z} \right ) ,
\label{EQw4.36}
\end{equation}
violates the conservation condition, Eq.~(\ref{EQw4.22}), for $\alpha=0$.
This result shows that the total energy and the total momentum being
constant in time does not guarantee that the evolution equations for the
total energy and the total momentum satisfy the local conservation law,
in fact, just the opposite.
\par
In an influential 1973 article, Gordon \cite{BIGord} uses a microscopic
model of the dielectric in terms of electric dipoles.
Assuming a dilute vapor in which the dipoles do not interact with each
other or their host, Gordon writes the microscopic Lorentz
dipole force on a particle with linear polarizability
$\alpha$ as \cite{BIAMC3,BIGord,BIPenHaus}
\begin{equation}
{\bf f}_{atom}=\alpha\left ( ({\bf e}\cdot\nabla){\bf e}
+ \frac{d{\bf e}}{dt}\times{\bf b}
\right ) 
\label{EQw4.37}
\end{equation}
in the vacuum,
where ${\bf e}$ is the microscopic electric field and ${\bf b}$
is the microscopic magnetic field.
The material momentum density is obtained by spatially averaging the
force on a single dipole and integrating with respect to time.
Then the material momentum density is
\begin{equation}
{\bf g}_{matl}=\int \langle N {\bf f}_{atom}\rangle dt \, ,
\label{EQw4.38}
\end{equation}
where $N$ is the dipole density.
The fields acting on the dipoles inside a dielectric are not the same
as the fields in free space.
For the purpose of presenting the prior work, we suffer, without proof,
that the material momentum density is \cite{BIGord}
\begin{equation}
{\bf g}_{matl}
= N\alpha \int \frac{{\bf E}\times{\bf B}}{c} dt
+N\alpha \int \frac{1}{2}\nabla ({\bf E}^2) dt \, .
\label{EQw4.39}
\end{equation}
Gordon assumes that the total momentum density is the sum of the
Abraham momentum density and the material momentum density. 
Making a transformation to retarded time \cite{BIIcsevgiLamb},
Gordon \cite{BIGord} derives
\begin{equation}
{\bf g}_G=\frac{n{\bf E}\times{\bf B}}{c} 
\label{EQw4.40}
\end{equation}
for the total momentum density and then assumes a pseudo-momentum 
in order to force agreement with the Minkowski form of momentum.
\par
In the Gordon model, and similar models, the dipoles are
free particles in the vacuum that are accelerated by the Lorentz dipole
force at the leading edge of the quasi-monochromatic field and travel at
constant velocity until decelerated by the Lorentz dipole force at the
trailing edge of the field.
In a real dielectric, or a more complete theoretical model of a
dielectric, the motion of the material dipoles will be considerably
impeded by collisions, lattice strains, or other effects of the host
material.
Consequently, it is assumed that a traveling deformation of the
material, rather than the unrestrained motion of dipoles, will
contribute the requisite material momentum \cite{BIPfei,BIGord}.
The Gordon linear momentum
\begin{equation}
{\bf G}_G=\int_{\Sigma}{\bf g}_G dv
=\int_{\Sigma}\frac{n{\bf E}\times{\bf B}}{c} dv
\label{EQw4.41}
\end{equation}
that is obtained by spatially integrating the Gordon
momentum density and the Minkowski momentum, Eq.~(\ref{EQw4.18}), that is
obtained by adding a hypothetical pseudomomentum concludes the derivation.
Comparing Eq.~(\ref{EQw4.41}) to Eq.~(\ref{EQw4.15}),
the Gordon momentum ${\bf G}_G$ is constant in time in the case of
propagation of a quasimonochromatic field through a gradient-index
antireflection-coated simple linear dielectric
\cite{BIBahder,BIJMP,BIGord}.
Then, we now have a plausible model for the material momentum. 
\par
There are several problems with the derivation presented in
Ref.~\cite{BIGord} in addition to the assumptions that are described
above:
\textit{i}) In Eq.~(\ref{EQw4.37}), the force density
$\alpha({\bf e}\cdot\nabla){\bf e}$ has been improperly retained because
several other terms of the same order have been dropped in the dipole
approximation \cite{BICrendrr}.
Moreover, this small term is divided into two large terms of nearly
equal magnitude and opposite sign and one of these terms is eliminated.
\textit{ii}) Temporal independence of the total linear momentum is only
one of the four conditions of the energy--momentum conservation laws,
Eqs.~(\ref{EQw3.07})--(\ref{EQw3.10}).
\textit{iii}) There is a factor of 2 error in the susceptibility used by
Gordon.
In the corrected version of the Gordon derivation, Milonni and Boyd
\cite{BIAMC3} prove that the sum of the electromagnetic and material
momentums is the Minkowski momentum, Eq.~(\ref{EQw1.02}), which is not
constant in time.
\par
Barnett and Loudon \cite{BIAMC2} and Barnett \cite{BIBarn} present a
model in which the Abraham momentum and the Minkowski momentum are both
appropriate momentums for the field in a dielectric.
Each of the classical electromagnetic momentums is accompanied by a
material momentum, different in each case, and identified with either
a canonical or kinetic phenomenology.
The material momentum densities, ${\bf g}_{canonical}^{matl}$ and
${\bf g}_{kinetic}^{matl}$, are defined implicitly by global conservation
of total momentum, such that
\begin{equation}
{\bf G}_{total}={\bf G}_M+\int_{\Sigma} {\bf g}_{canonical}^{matl} dv
={\bf G}_A+ \int_{\Sigma}{\bf g}_{kinetic}^{matl} dv \, ,
\label{EQw4.42}
\end{equation}
where ${\bf G}_{total}={\bf G}_{incident}$ for a gradient-index
anti-reflection coated simple linear dielectric block.
Although providing a descriptive model for construction of the total
linear momentum, the total linear momentum, ${\bf G}_{total}$, is
unique in a thermodynamically closed system because it is constant in
time and it is a known quantity in terms of macroscopic fields,
Eq.~(\ref{EQw4.23}).
As in the previous example, temporal independence of the total linear
momentum is only one of the four conditions of the spacetime
conservation laws, Eqs.~(\ref{EQw3.07})--(\ref{EQw3.10}).
If we use either of the canonical or kinetic models of
Eq.~(\ref{EQw4.42}) for the total linear momentum, then the total
energy--momentum tensor will be the same as Eq.~(\ref{EQw4.25}).
Applying the local conservation condition, Eq.~(\ref{EQw4.22}), the
four-divergence of the total energy--momentum tensor will produce
a demonstrably false energy continuity equation, just as before,
Eq.~(\ref{EQw4.26}).
Again, the two nonzero terms in Eq.~(\ref{EQw4.26}) depend on different
powers of the refractive index and Eq.~(\ref{EQw4.26}) is incommensurate
with the Poynting theorem.
\par
Ramos, Rubilar, and Obukhov \cite{BIObuk} utilize a fully relativistic 
4-dimensional tensor formalism to discuss the energy--momentum of a
system that consists of an antireflection-coated rigid slab of
dielectric with a final constant velocity ${\bf v}$.
Their total energy--momentum tensor is
$$
{\sf T}_{\mu}^{\;\; \nu}
=\rho_0 u_{\mu}u^{\nu}
+\Big ( {\sf F}_{\mu\sigma}F^{\sigma\nu}
+\frac{1}{4}
\delta_{\mu}^{\nu}F^{\sigma\lambda}{\sf F}_{\sigma\lambda}\Big )
$$ 
$$
+(n^2-1)
\Big ({\sf F}_{\mu\sigma}F^{\lambda\nu}u^{\sigma}u_{\lambda}
+\frac{1}{2}
\delta_{\mu}^{\nu}{\sf F}_{\sigma\rho_0}F^{\sigma\lambda}
u^{\rho_0}u_{\lambda}
$$
\begin{equation}
- F^{\rho_0\sigma}{\sf F}_{\rho_0\lambda}
u_{\sigma}u^{\lambda}u_{\mu}u^{\nu}\Big ) \, ,
\label{EQw4.43}
\end{equation}
where $u^{\mu}$ is the four-velocity $(\gamma,\gamma {\bf v})$.
Ramos, Rubilar, and Obukhov \cite{BIObuk} claim that the total
energy--momentum tensor, Eq.~(\ref{EQw4.43}), satisfies the
energy--momentum balance equation
\begin{equation}
\partial_{\nu}{\sf T}_{\mu}^{\;\; \nu}-{\sf F}_{\mu\nu}J^{\nu}_{ext}=0\, ,
\label{EQw4.44}
\end{equation}
that the energy--momentum tensor of the complete system is conserved,
and that the system is thermodynamically closed if the four-current
density $J^{\nu}_{ext}$ is zero.
Then the total four-momentum of the whole system is globally conserved and
\begin{equation}
P_j=\int_{\Sigma}{\sf T}_{j}^{\;\; 0} dv
\label{EQw4.45}
\end{equation}
is a time-independent quantity \cite{BIObuk}.
In order to test the validity of the total energy--momentum tensor,
Eq.~(\ref{EQw4.43}), we consider a quasimonochromatic field in the
plane-wave limit to be normally incident on a simple linear dielectric
through a gradient-index antireflection coating.
Evaluating the $\nu=0$ element of Eq.~(\ref{EQw4.44}), we obtain
\begin{equation}
\frac{\partial}{\partial (ct)}
\left [\rho_0 c^2+ \frac{1}{2}(n^2{\bf E}^2+{\bf B}^2) \right ]
+\nabla\cdot(\rho_0 {\bf v})
+\nabla\cdot ( {\bf E}\times{\bf B})=0
\label{EQw4.46}
\end{equation}
by substitution from the total energy--momentum tensor,
Eq.~(\ref{EQw4.43}), using quantities from the usual antisymmetric field
tensor
\begin{equation}
{\sf F}^{\alpha\beta}=
\left [
\begin{matrix}
0
\!&-E_x
\!&-E_y
\!&-E_z
\cr
E_x       &0               &-B_z            &B_y
\cr
E_y       &B_z             &0               &-B_x
\cr
E_z       &-B_y            &B_x             &0 
\cr
\end{matrix}
\right ] \, .
\label{EQw4.47}
\end{equation}
Now, Eq.~(\ref{EQw4.46}) is the same as Eq.~(\ref{EQw4.31}) that was
derived previously using the model of 
Pfeifer, Nieminen, Heckenberg, and Rubinsztein-Dunlop \cite{BIPfei}.
Substituting Eq.~(\ref{EQw4.32}) into Eq.~(\ref{EQw4.46}) and taking
$\rho_0$ as constant in time \cite{BIObuk}, we write
\begin{equation}
\frac{1}{c}\frac{\partial}{\partial t}
\left [ \frac{1}{2}\left (n^2{\bf E}^2+{\bf B}^2\right )\right ]
+\nabla \cdot \left (n{\bf E}\times {\bf B} \right ) =0\, .
\label{EQw4.48}
\end{equation}
As before, Eq.~(\ref{EQw4.48}) {\it i}) is incommensurate with the
Poynting theorem and {\it ii}) is self-inconsistent because the two
non-zero terms depend on different powers of the refractive index
thereby disproving the energy--momentum balance equation,
Eq.~(\ref{EQw4.44}), with $J^{\nu}_{ext}=0$.
\par
A fully microscopic model of the interaction of light with ponderable
matter is unique, valid, and beyond our capabilities.
The examples above are representative of the many diverse
quasi-microscopic treatments of the Abraham--Minkowski controversy and
there is no unique quasi-microscopic model \cite{BIPfei}.
There are many ways to couple and average the quasi-microscopic material
properties with the electromagnetic properties that are systematically
derived from the macroscopic Maxwell--Minkowski equations.
The correctness of the procedure is rooted in the fundamental basis of
the model and the derivation.
In the end, the total linear momentum is the sum of the electromagnetic
momentum and the material momentum.
However, the spacetime conservation laws are not satisfied.
We provided specific examples, using the models of 
Pfeifer, Nieminen, Heckenberg, and Rubinsztein-Dunlop \cite{BIPfei},
of Barnett \cite{BIBarn}, and of
Ramos, Rubilar, and Obukhov \cite{BIObuk}, where the continuity equation
for the total energy is proven to be false.
More importantly, these are general results as shown in Sec. 4B.
Any construction of the total energy--momentum tensor must be based on
energy and momentum densities corresponding to the time
independent total energy, Eq.~(\ref{EQw4.24}) and the time
independent total momentum, Eq.~(\ref{EQw4.23}).
Then \textit{the four-divergence of the total energy--momentum tensor
that is constructed using Maxwellian continuum electrodynamics will 
always result in a provably false energy continuity equation}, even if
a phenomenological material energy--momentum tensor is assumed.
\par
\section{Lagrangian Field Dynamics in a Dielectric-Filled Spacetime}
\par
At the fundamental microscopic level, dielectrics consist of tiny
bits of host and polarizable matter, embedded in the vacuum, with
interactions of various types.
According to Lorentz, the seat of the electromagnetic field is 
empty space.
If a light pulse is emitted from a point $(x_a,y_a,z_a)$ at time $t_a$
then spherical wavefronts are defined by
\begin{equation}
(x-x_a)^2+(y-y_a)^2+(z-z_a)^2=\left ( c(t-t_a) \right )^2
\label{EQw5.01}
\end{equation}
in a flat four-dimensional Minkowski spacetime $S_v(ct,x,y,z)$.
Equation (\ref{EQw5.01}) underlies classical electrodynamics and its
relationship to special relativity.
Although light always travels at speed $c$ \cite{BIFeynman},
Eq.~(\ref{EQw5.01}) is only valid at very short range before the light
is scattered by the various microscopic features of the dielectric.
While the microscopic picture is always valid, there are practical 
difficulties in treating all of the interactions as light traverses
a dielectric.
\par
In continuum electrodynamics, the dielectric is treated as continuous
at all length scales and the macroscopic refractive index $n$ is defined
such that light travels with an effective speed of $c/n$.
In an arbitrarily large simple linear dielectric medium with an
isotropic homogeneous index of refraction $n$, spherical wavefronts from
a point source at $(x_a,y_a,z_a)$ and emitted at time $t_a$ are defined
by
\begin{equation}
(x-x_a)^2+(y-y_a)^2+(z-z_a)^2=\left ( \frac{c(t-t_a)}{n} \right )^2 \, .
\label{EQw5.02}
\end{equation}
At this point, we postulate Eq.~(\ref{EQw5.02}), instead of 
Eq.~(\ref{EQw5.01}), as the basis of a theory of continuum electrodynamics
and derive the consequences for field theory, classical continuum
electrodynamics, special relativity, spacetime, and experiments.
\par
We consider an arbitrarily large region of space to be filled with a
simple linear isotropic homogeneous dielectric that is characterized by
a linear refractive index $n$.
For clarity and concision, we will work in a regime in which dispersion
can be treated parametrically and is otherwise negligible such that
$n(\omega_p)$ is a real time-independent constant for a transparent
dielectric that is illuminated by a quasimonochromatic field of center
frequency $\omega_p$, as described in Sec.~2.
We define an inertial reference frame $S(x,y,z)$ with orthogonal
axes, $x$, $y$, and $z$, and require that the origin of the reference
frame is significantly inside the volume that is defined by the surface
of the dielectric medium.
We denote a time-like coordinate in the medium as $\bar x_0$.
If a light pulse is emitted from the origin at time
\begin{equation}
\tau=\bar x_0/c=0 \, ,
\label{EQw5.03}
\end{equation}
then
\begin{equation}
x^2+y^2+z^2=\left ( \bar x_0 \right )^2
\label{EQw5.04}
\end{equation}
describes spherical wavefronts in $S$.
The four-vector $(\bar x_0,x,y,z)$ represents the
position of a point in a four-dimensional, isotropic, homogeneous, flat,
non-Minkowski material spacetime $S_d(\bar x_0,x,y,z)$.
Clearly, the new time-like coordinate $\bar x_0$ is associated with
$c\tau=ct/n$, while $x_0=c t$ is the usual time-like coordinate
in a vacuum Minkowski spacetime.
The material spacetime reduces to ordinary Minkowski spacetime if $n=1$.
\par
The basis functions,
$\exp(-i(n\omega/c)(\bar x_0-{\bf \hat k}_0\cdot{\bf r}))$,
define the null surface, $\bar x_0={\bf \hat k}_0\cdot{\bf r}$.
Fig.~3 is a depiction of the intersection of the light cone with the
$x-\bar x_0$ plane in the flat material spacetime $S_d$ showing the
null $\bar x_0= x$.
There will be a different material spacetime for each value of the 
refractive index, but the half-opening angle of the
material light cone will always be $\alpha=\pi/4$ in the corresponding
material spacetime.
The unit slope of the null in the $x-\bar x_0$ plane of the
non-Minkowski material spacetime is related to the coordinate speed of
light in a simple linear dielectric by
\begin{equation}
\frac{\Delta x}{\Delta t}=
\frac{\Delta x}{\Delta \bar x_0}
\frac{d \bar x_0}{dt} = 1\cdot\frac{c}{n} =\frac{c}{n} \, .
\label{EQw5.05}
\end{equation}
This equation shows that the effective speed of light in a
simple linear dielectric medium is attributable to 
renormalization of the time-like coordinate by $n$. 
\begin{figure}
\includegraphics[scale=0.22]{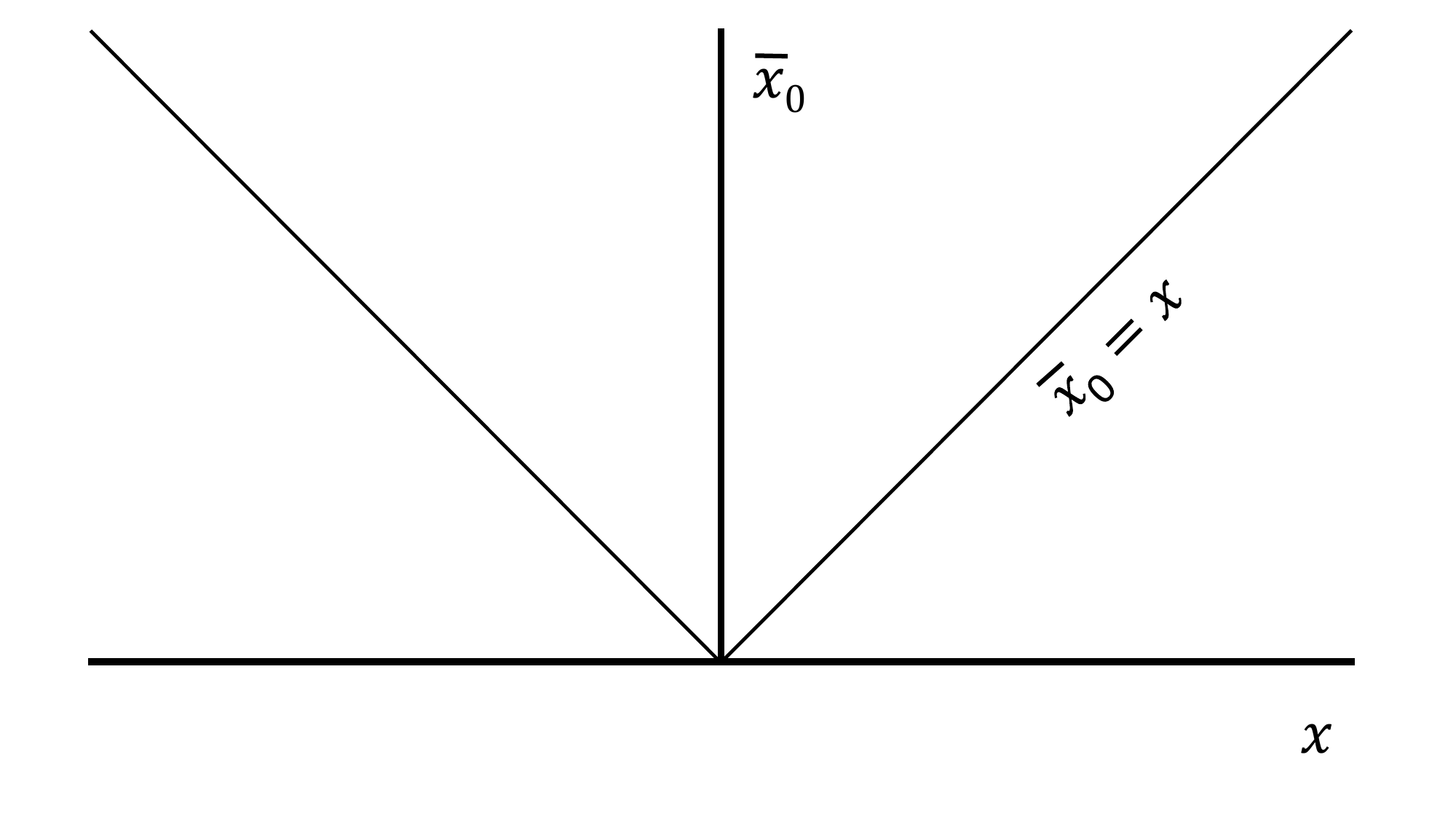}
\caption{Null cone for light depicted in the $\bar x_0-x$ plane
of a flat, non-Minkowski material spacetime that corresponds to a
simple linear dielectric.}
\label{fieldfig3}
\end{figure}
\par
For a system of particles, the transformation of the position
vector ${\bf x}_i$ of the $i^{th}$ particle to $J$ independent
generalized coordinates is
\begin{equation}
{\bf x}_i={\bf x}_i(\tau;q_1,q_2, \ldots, q_J) .
\label{EQw5.06}
\end{equation}
Applying the chain rule, we obtain the virtual displacement
\begin{equation}
\delta{\bf x}_i=\sum_{j=1}^J
\frac{\partial {\bf x}_i}{\partial q_j}\delta q_j
\label{EQw5.07}
\end{equation}
and the velocity
\begin{equation}
{\bf u}_i=\frac{d{\bf x}_i}{d\tau}=
\sum_{j=1}^J
\frac{\partial {\bf x}_i}{\partial q_j}
\frac{d q_j}{d \tau}
+ \frac{\partial {\bf x}_i}{\partial \tau}
\label{EQw5.08}
\end{equation}
of the $i^{th}$ particle in the new coordinate system.
Substitution of
\begin{equation}
\frac{\partial{\bf u}_i}{\partial(d q_j/d\tau)}=
\frac{\partial {\bf x}_i}{\partial q_j}
\label{EQw5.09}
\end{equation}
into the identity
\begin{equation}
\frac{d}{d\tau}\left ( m{\bf u}_i\cdot
\frac{\partial{\bf x}_i}{\partial q_j} \right ) =
m\frac{d{\bf u}_i}{d \tau}\cdot
\frac{\partial {\bf x}_i}{\partial q_j}
+
m{\bf u}_i\cdot\frac{d}{d\tau}
\left ( \frac{\partial{\bf x}_i}{\partial q_j}\right )
\label{EQw5.10}
\end{equation}
yields
\begin{equation}
\frac{d{\bf p}_i}{d\tau}\cdot
\frac{\partial{\bf x}_i}{\partial q_j} =
\frac{d}{d\tau}
\left ( \frac{\partial}{\partial(d q_j/d \tau)}
\frac{1}{2}m{\bf u}_i^2
\right ) -
\frac{\partial}{\partial q_j}\left ( \frac{1}{2}m{\bf u}_i^2\right ).
\label{EQw5.11}
\end{equation}
\par
For a system of particles in equilibrium, the virtual work of the
applied forces ${\bf f}_i$ vanishes and the virtual work on each
particle vanishes leading to the principle of virtual work
\begin{equation}
\sum_i{\bf f}_i\cdot \delta{\bf x}_i=0
\label{EQw5.12}
\end{equation}
and D'Alembert's principle
\begin{equation}
\sum_i\left ( {\bf f}_i -\frac{d{\bf p}_i}{d\tau}\right )
\cdot \delta{\bf x}_i=0.
\label{EQw5.13}
\end{equation}
Using Eqs.~(\ref{EQw5.07}) and (\ref{EQw5.11})
and the kinetic energy of the $i^{th}$
particle
\begin{equation}
{T}_i= \frac{1}{2} m{\bf u}_i^2,
\label{EQw5.14}
\end{equation}
we can write D'Alembert's principle, Eq.~(\ref{EQw5.13}), as
\begin{equation}
\sum_j^J \left [ \left ( \frac{d}{d\tau}
\left (
\frac{\partial T}{\partial(d q_j/d \tau)}\right )
-\frac{\partial T}{\partial q_j}
\right ) -Q_j \right ] \delta q_j =0
\label{EQw5.15}
\end{equation}
in terms of the generalized forces
\begin{equation}
Q_j=\sum_i{\bf f}_i\cdot
\frac{\partial{\bf x}_i}{\partial q_j}.
\label{EQw5.16}
\end{equation}
If the generalized forces come from a generalized scalar
potential function $V$ \cite{BIGold}, then we can write
Lagrange equations of motion
\begin{equation}
\frac{d}{d\tau} \left (
\frac{\partial L}{\partial(\partial q_j/\partial \tau)}\right )
- \frac{\partial L}{\partial q_j} =0,
\label{EQw5.17}
\end{equation}
where $L=T-V$ is the Lagrangian.
The canonical momentum is therefore
\begin{equation}
p_j=\frac{\partial L}{\partial(\partial q_j/\partial \tau)}
\label{EQw5.18}
\end{equation}
in a linear medium.
Comparable derivations for the vacuum case, $\tau \rightarrow t$, appear
in Goldstein \cite{BIGold} and Marion \cite{BIMar}, for example.
This version of canonical momentum differs from the existing
vacuum formula because the material time $\tau$ appears instead
of the vacuum time $t$.
\par
The field theory \cite{BICT} is based on a generalization of
the discrete case in which the dynamics are derived from a Lagrangian
density ${\cal L}$.
The generalization of the Lagrange equation, Eq.~(\ref{EQw5.17}),
for fields in a linear medium is
\begin{equation}
\frac{d}{d \bar x_0}\frac{\partial{\cal L}}
{\partial (\partial A_j /\partial \bar x_0)}
=\frac{\partial {\cal L}}{\partial A_j}
-\sum_i\partial_{i}
\frac{\partial{\cal L}}{\partial(\partial_{i} A_j )} \, .
\label{EQw5.19}
\end{equation}
This equation differs from the Lagrange equation
for fields in the vacuum \cite{BICT}
\begin{equation}
\frac{d}{d x_0}\frac{\partial{\cal L}}
{\partial (\partial A_j /\partial x_0)}
=\frac{\partial {\cal L}}{\partial A_j}
-\sum_i\partial_{i}
\frac{\partial{\cal L}}{\partial(\partial_{i} A_j )} 
\label{EQw5.20}
\end{equation}
in that differentiation is performed with respect to the material
time-like coordinate $\bar x_0$ instead of the vacuum coordinate $x_0$.
We take the Lagrangian density of the electromagnetic field in the
medium to be
\begin{equation}
{\cal L}=
\frac{1}{2}
\left ( \left (
\frac{\partial{\bf A}}{\partial \bar x_0} \right )^2
-(\nabla\times{\bf A})^2 \right ) \, .
\label{EQw5.21}
\end{equation}
Again, differentiation is performed with respect to the material
time-like coordinate $\bar x_0$ instead of the vacuum coordinate $x_0$.
Furthermore, the Lagrangian density is explicitly quadratic in
the macroscopic fields corresponding to real eigenvalues and a 
conservative system.
\par
Equations (\ref{EQw5.19}) and (\ref{EQw5.21}) form the basis for a new
canonical theory of macroscopic fields in a simple linear dielectric.
The new theory has similarities in appearance to the macroscopic Maxwell
equations, but it is disjoint from the Maxwell theory because it is
based in a flat non-Minkowski material spacetime $S_d(\bar x_0,x, y,z)$
instead of a vacuum Minkowski spacetime $S_v(x_0,x, y,z)$.
Constructing the components of Eq.~(\ref{EQw5.19}), we have
\begin{equation}
\frac{\partial{\cal L}}
{\partial (\partial A_{j}/\partial \bar x_0)}
=\frac{\partial A_j}{\partial \bar x_0}
\label{EQw5.22}
\end{equation}
\begin{equation}
\frac{\partial \cal L}{\partial A_j}= 0
\label{EQw5.23}
\end{equation}
\begin{equation}
\sum_i\partial_{i}
\frac{\partial{\cal L}}{\partial(\partial_{i} A_{j})}
=[\nabla\times(\nabla\times {\bf A})]_j
\label{EQw5.24}
\end{equation}
for the Lagrangian density given in Eq.~(\ref{EQw5.21}).
We substitute the individual terms,
Eqs.~(\ref{EQw5.22})--(\ref{EQw5.24}), into Eq.~(\ref{EQw5.19}), then
the Lagrange equations of motion for the electromagnetic field in a
dielectric are the three orthogonal components of the vector wave
equation
\begin{equation}
\nabla\times(\nabla\times {\bf A})
+ \frac{\partial^2{\bf A}}{\partial \bar x_0^2}
=0 \, .
\label{EQw5.25}
\end{equation}
For fields, the canonical momentum density
\begin{equation}
\Pi_j= \frac{\partial{\cal L}}
{\partial (\partial A_{j}/\partial \bar x_0)}
\label{EQw5.26}
\end{equation}
from Eq.~(\ref{EQw5.22}) supplants the discrete canonical
momentum defined in Eq.~(\ref{EQw5.18}).
\par
We can write the second-order equation, Eq.~(\ref{EQw5.25}), as a
set of first-order differential equations.
To that end, we introduce macroscopic field variables
\begin{equation}
{\bf \Pi}=
\frac{\partial{\bf A}}{\partial \bar x_0}
\label{EQw5.27}
\end{equation}
\begin{equation}
{\bf B}=\nabla\times{\bf A}
\label{EQw5.28}.
\end{equation}
Here, ${\bf \Pi}$ is the canonical momentum field density whose
components were defined in Eq.~(\ref{EQw5.26}) after making the
substitutions indicated by Eq.~(\ref{EQw5.22}).
Substituting the definition of the canonical momentum field ${\bf \Pi}$,
Eq.~(\ref{EQw5.27}), and the definition of the magnetic field ${\bf B}$,
Eq.~(\ref{EQw5.28}), into Eq.~(\ref{EQw5.25}),
we obtain the Maxwell--Amp\`ere-like law
\begin{equation}
\nabla\times{\bf B}+\frac{\partial{\bf \Pi}}{\partial \bar x_0}
=0 \,.
\label{EQw5.29}
\end{equation}
The divergence of Eq.~(\ref{EQw5.28}) and the curl of
Eq.~(\ref{EQw5.27}) respectively produce Thompson's Law
\begin{equation}
\nabla\cdot{\bf B}=0
\label{EQw5.30}
\end{equation}
and a Faraday-like law
\begin{equation}
\nabla\times{\bf \Pi}
- \frac{\partial{\bf B}}{\partial \bar x_0}
= \frac{\nabla n}{n}\times{\bf \Pi} \, .
\label{EQw5.31}
\end{equation}
The divergence of the variant Maxwell--Amp\`ere Law,
Eq.~(\ref{EQw5.29}), is
\begin{equation}
\frac{\partial}{\partial \bar x_0}\nabla\cdot{\bf \Pi}=
-\frac{\nabla n}{n}\cdot\frac{\partial {\bf \Pi}}{\partial \bar x_0}\, .
\label{EQw5.32}
\end{equation}
Integrating Eq.~(\ref{EQw5.32}) with respect to the time-like coordinate
yields a modified version of Gauss's law
\begin{equation}
\nabla\cdot{\bf \Pi}=-\frac{\nabla n}{n}\cdot {\bf \Pi} -c_1
\label{EQw5.33}
\end{equation}
where $-c_1$ is a constant of integration.
Based on the derivation of these equations, it is required that the 
source terms in Eqs.~(\ref{EQw5.31}) and (\ref{EQw5.33}) that
involve the gradient of the refractive index are, at most, perturbative,
essentially limiting the theory to an isotropic homogeneous block of
simple linear dielectric draped with a gradient-index antireflection
coating or a piecewise homogeneous simple linear dielectric.
We have not included free charges and a free-charge current because it is
an unnecessary complication and because an inviscid incoherent
flow of non-interacting charges in the continuum limit moving unimpeded
through a continuous dielectric cannot be justified at the level of
rigor that we are employing in the current work.
\par
This completes the set of first-order equations of motion for the
macroscopic fields, Eqs.~(\ref{EQw5.29})--(\ref{EQw5.31}) and
(\ref{EQw5.33}).
Consolidating the equations of motion and dropping the inhomogeneous
source terms, we have the equations of motion
for macroscopic electromagnetic fields in an isotropic homogeneous
simple linear dielectric,
\begin{subequations}
\begin{equation}
\nabla\times{\bf B}+\frac{\partial{\bf \Pi}}{\partial \bar x_0}=0
\label{EQw5.34a}
\end{equation}
\begin{equation}
\nabla\cdot{\bf B}=0
\label{EQw5.34b}
\end{equation}
\begin{equation}
\nabla\times{\bf \Pi}-\frac{\partial{\bf B}}{\partial \bar x_0}=0
\label{EQw5.34c}
\end{equation}
\begin{equation}
\nabla\cdot{\bf \Pi}= 0 \, ,
\label{EQw5.34d}
\end{equation}
\label{EQw5.34}
\end{subequations}
derived from field theory for quasimonochromatic electromagnetic
fields in a linear dielectric-filled, flat, non-Minkowski continuous
material spacetime.
\par
Readers of this article may note that the macroscopic field equations,
Eqs.~(\ref{EQw5.34}), obviously violate special relativity because
${\bar x}_0$ is index-dependent.
It is also obvious that Eqs.~(\ref{EQw5.34}) cannot possibly violate
special relativity because their form is isomorphic to an easily verified
identity of the macroscopic Maxwell--Minkowski equations,
Eqs.~(\ref{EQw2.01}), which are known to satisfy special relativity.
Readers should not unequivocally, uncritically, and untenably
advocate one side or the other of this contradiction based solely on the 
appearance of Eqs.~(\ref{EQw5.34}).
It is shown in Ref.~\cite{BIPrepr} that Eqs.~(\ref{EQw5.34}) comply with
Rosen dielectric special relativity.
The applicability of the Fresnel relations to the system described by 
Eqs.~(\ref{EQw5.34}) has been questioned and treated in Ref.~\cite{BIFresnel}.
Other issues like free charges, free charge currents, experimental verification,
dispersion, plane-wave limit, material motion, etc. were addressed as they came up in the description.
\par
We can never place a matter-based observer, no matter how small, in a
continuous dielectric because the model dielectric is continuous at all
length scales and will always be displaced.
Consequently, the necessity to make non-optical measurements in a vacuum
leads to the establishment of a Laboratory Frame of Reference.
An observer that resides in the vacuum by virtue of displacing the
terrestrial atmosphere outside of the dielectric, such as
Fizeau \cite{BIFizeau}, will measure the speed of light in a dielectric to
be dependent on the velocity of the dielectric relative to the vacuum-based
Laboratory Frame of Reference, Eq.~(\ref{EQw2.11}), an effect that Fresnel
attributed to ether drag.
\par
In deriving the theory of dielectric special relativity, Laue considered
a block of dielectric moving inertially in a Laboratory Frame of Reference
and used the Einstein velocity sum rule to derive the theory of dielectric
special relativity and confirm Fresnel drag.
That physical configuration is not correct for the special relativity that
underlies the equations of motion for electromagnetic fields in a
dielectric, Eqs.~(\ref{EQw5.34}).
The current author \cite{BIPrepr} derived a theory of dielectric special
relativity for inertial reference frames translating at constant speed in
an arbitrarily large region of space in which the speed of light is $c/n$
in the local rest frame.
This rigorous derivation of special relativity in an appropriate physical
configuration confirms Rosen's \cite{BIRosen} phenomenological derivation
of an index-dependent theory of special relativity in a dielectric.
The speed of light at the location of the observer in the dielectric is
obviously independent of the motion of the dielectric; otherwise there
would be an inaccessible preferred Laboratory Reference Frame.
The situation is different from the Laue theory; there the vacuum-based
Laboratory Frame of Reference is established first and the motion of the
dielectric in the preferred Laboratory Reference Frame can be specified and
measured.
\par
Rosen \cite{BIRosen} noted that there will be a different theory of
relativity associated with a limiting speed in each material.
In the rest frame of the material, the speed of light $c_d$ will be
different in different dielectric materials and we can label different
materials with the index $i$.
Considering only isotropic, homogeneous linear dielectric materials in
which the speed of light is inversely proportional to a real constant
$n_i$, we obtain
\begin{equation}
({\gamma_d})_i= \frac {1}{\sqrt{1- n_i^2v_d^2/c^2}}
\label{EQw5.35}
\end{equation}
as our material-specific Lorentz factor \cite{BIPrepr,BIRosen}.
As discussed above, the configuration of the physical system that we
are treating is different from the system that was employed by Laue
\cite{BILaue,BILaue2}.
\par
In this article, the theory of quasimonochromatic radiation interacting
with a simple linear dielectric has been discussed primarily in terms
of an arbitrarily large isotropic homogeneous medium or a block of an
isotropic, homogeneous, linear dielectric material draped with a
gradient-index anti-reflection coating.
At some point, we will be required to deal with the boundary conditions
of piecewise homogeneous linear dielectric materials.
Reflection and refraction are experimentally uncomplicated and it would
be unpleasant if the usual Fresnel formulas failed to work for the new
theory.
On the other hand, it is apparent that the usual derivation of the
Fresnel relations \cite{BIMar,BIGriff,BIJackson,BIZangwill} by
application of the Stoke's theorem and the divergence theorem to the
Maxwell--Minkowski equations will not work when applied to the new field
equations, Eqs.~(\ref{EQw5.34}).
Boundary conditions and the Fresnel relations are rigorously derived
by conservation of energy and the application of Stoke's theorem to the
wave equation in a separate publication \cite{BIFresnel}.
\par
We cannot rigorously relate the ordinary macroscopic Maxwell--Minkowski
field equations, Eqs.~(\ref{EQw2.01}), to the results of our derivation,
Eqs.~(\ref{EQw5.34}).
In the current formalism, based on Lagrangian field theory adapted to
a region of space in which the speed of light is $c/n$, the usual
macroscopic Maxwell fields ${\bf E}$ and ${\bf D}$ are not definable in
terms of ${\bf \Pi}$ because the refractive index is contained in the
independent coordinate and is not a free material parameter.
As disclosed by Rosen \cite{BIRosen}, there is a different theory of
relativity associated with each isotropic homogeneous medium in which a
limiting speed is associated with the phenomena that take place in the
medium.
Likewise, there is a different theory of electrodynamics for each linear
medium, labelled $i$, with refractive index $n_i$ and the different
theories correspond to disjoint isotropic, homogeneous, flat,
non-Minkowski material spacetimes \cite{BIRosen}
$(S_d)_i(\bar x_{0_{ii}},x,y,z)$.
Then the macroscopic Maxwell--Minkowski equations, Eqs.~(\ref{EQw2.01}),
cannot be derived as valid theorems from our new axioms,
Eqs.~(\ref{EQw5.34}), and we cannot insert, by hand, familiar expressions
that were derived from Maxwellian continuum electrodynamics into the new
theory.
\par
Nevertheless, we would like to informally compare and contrast the new
theory with Maxwellian continuum electrodynamics.
There is sufficient commonality with the classic theories that 
simple phenomena like refraction and wave velocity
have equivalent formulations in the new theory.
Then, there are special cases in which the new
\{${\bf \Pi}$,${\bf B}$\} continuum electrodynamics can be
phenomenologically related to the usual Minkowski representation
\{${\bf E}$,${\bf B}$,${\bf D}$,${\bf H}$\} of the macroscopic Maxwell
equations.
These typically involve situations in which only one of the
equations of macroscopic electrodynamics is needed, such as the wave
equation
\begin{equation}
\nabla\times(\nabla\times {\bf A})
+ \frac{\partial}{\partial \bar x_0}
\left ( \frac{\partial {\bf A}}{\partial \bar x_0}\right )=0 
\label{EQw5.36}
\end{equation}
that is derived by substituting the definitions of the macroscopic
fields, Eqs.~(\ref{EQw5.27}) and (\ref{EQw5.28}), into
the Maxwell--Amp\`ere-like law, Eq.~(\ref{EQw5.34a}).
Therefore, we can be assured that the extensive theoretical and
experimental work that is ``correctly'' described by the macroscopic
Maxwell theory of continuum electrodynamics has an equivalent,
or nearly equivalent, expression in the new theory.
Nevertheless, we must be very careful about integrating established
concepts and formulas of Maxwellian electrodynamics into the new version of
continuum electrodynamics.
\par
More interesting is the work that we can do with the new formalism of
continuum electrodynamics that was improperly posed in the standard
Maxwell theory of continuum electrodynamics.
These cases will typically involve the invariance or tensor properties
of the set of coupled equations of motion.
This interpretation is borne out in our common experience: the
macroscopic Maxwell--Minkowski equations produce exceedingly accurate
experimentally verified predictions of simple phenomena like reflection,
refraction, Fresnel relations, wave propagation, etc., but fail to
render a unique, uncontroversial, experimentally verifiable prediction
of energy--momentum conservation.
In the next section, we will demonstrate the utility of the new
formalism of continuum electrodynamics by addressing energy--momentum
conservation in a dielectric.
\par
\section{Conservation Laws and $\{{\bf \Pi},{\bf B}\}$ Electrodynamics}
\par
The derivation of the continuity equation of a property flux density
is described in Sec. 3 as applying the divergence
theorem to a Taylor series expansion of the property density
field $\rho$ and the property flux density field ${\bf g}=\rho{\bf u}$
to a continuous flow in an otherwise empty volume \cite{BIFox}.
Here, we are treating the continuous (continuum limit) flow of photons
(light field) in an arbitrarily large, isotropic, homogeneous, simple
linear dielectric that is modeled as a region of space in which the 
speed of light is $c/n$.
Therefore, the conditions on the flow differ from the vacuum conditions
that are assumed for the usual spacetime conservation laws that were
discussed in Sec.~3.
Microscopically, dielectrics are mostly empty space.
But, in the continuum limit, dielectrics are continuous at all
length scales and the light field cannot be treated as if it is flowing
in an otherwise empty volume.
\par
It is necessary to modify the spacetime conservation laws that were
presented in Sec.~3 for the flow of a photon fluid (light field) in a
non-empty volume.
As shown in Eq.~(\ref{EQw5.06}), the generalized temporal coordinate
in a dielectric-filled volume is $\tau$.
Then a continuity equation has the form
\begin{equation}
\frac{\partial\rho}{\partial \tau}+\nabla\cdot (\rho {\bf u})=0
\label{EQw6.01}
\end{equation}
in an arbitrarily large region of space that is filled with an
isotropic, homogeneous, simple linear dielectric material.
\par
We can compare the conservation laws in a dielectric-filled spacetime
with the vacuum conservation laws 
Eqs.~(\ref{EQw3.07})--(\ref{EQw3.10}), in an empty volume.
\par
1) Continuity equations in a dielectric have the form of
Eq.~(\ref{EQw6.01}), instead of Eq.~(\ref{EQw3.01}).
Writing a scalar continuity equation for energy and a scalar continuity
equation for each of the three components of the momentum, row-wise, we
obtain the differential equation
\begin{equation}
\bar\partial_{\beta}{\sf T}^{\alpha\beta}=0
\label{EQw6.02}
\end{equation}
instead of Eq.~(\ref{EQw3.03}) as a condition for conservation of 
energy and momentum for a continuous unimpeded flow in a 
dielectric-filled spacetime.
The \textit{material} four-divergence operator
\begin{equation}
\bar\partial_{\beta}=
\left (\frac{1}{c} \frac{\partial}{\partial \tau},
\frac{\partial}{\partial x},\frac{\partial}{\partial y},
\frac{\partial}{\partial z} \right ) 
\label{EQw6.03}
\end{equation}
replaces Eq.~(\ref{EQw2.29}) because $\tau$, not $t$, is the 
independent temporal coordinate.
\par
2) The total energy and the total linear momentum 
\begin{equation}
P^{\alpha}(\tau)=
\int_{\Sigma} {\sf T}^{\alpha 0} dv =
P^{\alpha}(\tau_0)
\label{EQw6.04}
\end{equation}
are constant in material time $\tau$ for each $\alpha$ (global conservation).
\par
3) The trace of the total energy--momentum tensor is proportional to the
mass density $\rho_m$
\begin{equation}
{{\sf T}^{\alpha \alpha}} \propto \rho_m 
\label{EQw6.05}
\end{equation}
and is zero for light.
\par
4) If the total energy--momentum tensor of the incident field is
diagonally symmetric then the total energy--momentum tensor 
inside the dielectric medium is most-likely diagonally symmetric
\begin{equation}
{\sf T}^{\alpha \beta}= {\sf T}^{\beta\alpha } 
\label{EQw6.06}
\end{equation}
as a matter of conservation of total angular momentum, absent
pathological boundary conditions, subsystem separation, or other
inappropriate system definitions.
\par
5) The extra continuity equation for the total energy and the total 
momentum in a thermodynamically closed system
\begin{equation}
\bar\partial_{\beta}{\sf T}^{\beta\alpha}=0
\label{EQw6.07}
\end{equation}
is obtained by substituting the symmetry condition,
Eq.~(\ref{EQw6.06}), into the continuity
condition, Eq.~(\ref{EQw6.02}).
\par
For each different medium, there is a different material four-divergence
operator, Eq.~(\ref{EQw6.03}), and a different material four-continuity
equation, Eq.~(\ref{EQw6.02}), due to the dependence of the time-like
coordinate on the refractive index $n$.
\par
We can demonstrate consistency of the new formulation of continuum
electrodynamics with the conservation laws inside a dielectric-filled
spacetime.
The equations of motion for the macroscopic fields,
Eqs.~(\ref{EQw5.29}) and (\ref{EQw5.31}),
can be combined in the usual manner, using algebra and calculus,
to write an energy continuity equation
\begin{equation}
\frac{\partial }{\partial \bar x_0} \left [
\frac{1}{2}\left ( {\bf \Pi}^2+{\bf B}^2\right )
 \right ]+
\nabla\cdot \left ( {\bf B}\times{\bf \Pi}\right )
=\frac{\nabla n}{n} \cdot({\bf B}\times{\bf \Pi})
\label{EQw6.08}
\end{equation}
in terms of an electromagnetic energy density
\begin{equation}
\rho=\frac{1}{2}\left ( {\bf \Pi}^2+{\bf B}^2\right ) \, ,
\label{EQw6.09}
\end{equation}
an electromagnetic momentum density
\begin{equation}
{\bf g}=\frac{{\bf B}\times{\bf \Pi}}{c} \, ,
\label{EQw6.10}
\end{equation}
and a perturbative transient power density
\begin{equation}
p=\frac{\nabla n}{n} \cdot({\bf B}\times{\bf \Pi}) \, .
\label{EQw6.11}
\end{equation}
Likewise, we can combine Eqs.~(\ref{EQw5.29})--(\ref{EQw5.31}) and
(\ref{EQw5.33}) to derive the momentum continuity equation
(see Sec.~6.8 of Jackson, Ref.~\cite{BIJackson})
\begin{equation}
\frac{\partial}{\partial x_0} ({\bf B}\times{\bf \Pi})_i
+\sum_j\frac{\partial}{\partial x_j}{{\sf W}_T}_{ij}
={\bf f}_i \, ,
\label{EQw6.12}
\end{equation}
where the stress-tensor ${\sf W}_T$ is
\begin{equation}
{{\sf W}_T}_{ij}=-\Pi_i\Pi_j-B_iB_j+
\frac{1}{2} ({\bf \Pi}^2+{\bf B}^2)\delta_{ij} 
\label{EQw6.13}
\end{equation}
and 
\begin{equation}
{\bf f}={\bf \Pi}\times\left ( \frac{\nabla n}{n}\times{\bf \Pi}\right )
-\left ( \frac{\nabla n}{n}\cdot{\bf \Pi} \right ){\bf \Pi}
\label{EQw6.14}
\end{equation}
is a (gradient) force density.
Then the continuity equations, Eqs.~(\ref{EQw6.08}) and (\ref{EQw6.12}),
can be written, row-wise, as a differential equation
\begin{equation}
\bar\partial_{\beta}{\sf T}^{\alpha\beta}=(p,{\bf f}) \, ,
\label{EQw6.15}
\end{equation}
where 
\begin{equation}
{\sf T}^{\alpha\beta}=
\left [
\begin{matrix}
({\bf \Pi}^2+{\bf B}^2)/2 &({\bf B}\times{\bf \Pi})_1
&({\bf B}\times{\bf \Pi})_2 &({\bf B}\times{\bf \Pi})_3
\cr
({\bf B}\times{\bf \Pi})_1   &{{\sf W}_T}_{11}   &{{\sf W}_T}_{12}      &{{\sf W}_T}_{13}
\cr
({\bf B}\times{\bf \Pi})_2   &{{\sf W}_T}_{21}   &{{\sf W}_T}_{22}      &{{\sf W}_T}_{23}
\cr
({\bf B}\times{\bf \Pi})_3   &{{\sf W}_T}_{31}   &{{\sf W}_T}_{32}      &{{\sf W}_T}_{34}
\cr
\end{matrix}
\right ]
\label{EQw6.16}
\end{equation}
is an electromagnetic matrix and 
$$
f^{\alpha}= (p,{\bf f}) = \Bigg (
\frac{\nabla n}{n} \cdot ({\bf B}\times{\bf \Pi}),
$$
\begin{equation}
{\bf \Pi}\times\left ( \frac{\nabla n}{n}\times{\bf \Pi} \right )
-\left ( \frac{\nabla n}{n}\cdot{\bf \Pi} \right ){\bf \Pi}
\Bigg )
\label{EQw6.17}
\end{equation}
is the four-force density.
\par
We integrate Eq.~(\ref{EQw6.10}) over all-space $\Sigma$ and obtain
\begin{equation}
{\bf G}=\int_{\Sigma}\frac{{\bf B}\times{\bf \Pi}}{c} dv. 
\label{EQw6.18}
\end{equation}
Likewise, 
\begin{equation}
U=\int_{\Sigma}\frac{{\bf \Pi}^2+{\bf B}^2}{2} dv
\label{EQw6.19}
\end{equation}
is obtained by integrating Eq.~(\ref{EQw6.09}).
\par
We can apply a gradient-index antireflection coating to an isotropic
homogeneous simple linear dielectric in order to greatly suppress
reflections.
Analysis of the wave equation for a quasimonochromatic pulse entering
an antireflection coated simple linear dielectric from the vacuum shows
that the amplitude of the vector potential is reduced by $\sqrt{n}$ and
the width is reduced by $n$, Sec.~4.
Then the definitions of the macroscopic canonical field ${\bf \Pi}$ and
the macroscopic magnetic field ${\bf B}$, Eqs.~(\ref{EQw5.27}) and
(\ref{EQw5.28}), show that the macroscopic fields in the dielectric are
each greater than the incident vacuum fields by a factor of $\sqrt{n}$
compensating for the reduced width of the field in the dielectric.
Neglecting the small gradients, the electromagnetic momentum,
Eq.~(\ref{EQw6.18}), and the electromagnetic energy,
Eq.~(\ref{EQw6.19}), are conserved.
Then the electromagnetic momentum, Eq.~(\ref{EQw6.18}), is the total
momentum and the electromagnetic energy, Eq.~(\ref{EQw6.19}), is the total
energy.
Consequently, there is no significant energy or momentum contained in 
any hypothetical unobservable material subsystem and there is no need
for a mechanism in the theory to couple to any subsystem by a source or
sink of either energy or momentum.
In this limit, the right-hand side of Eq.~(\ref{EQw6.15}) is negligible.
Then
\begin{equation}
\bar\partial_{\beta}{\sf T}^{\alpha\beta}=0
\label{EQw6.20}
\end{equation}
conforms to the corrected spacetime conservation condition,
Eq.~(\ref{EQw6.02}).
The other conservation conditions,
Eqs.~(\ref{EQw6.04})--(\ref{EQw6.06}), are also satisfied.
Therefore, the macroscopic electromagnetic system, Eqs.~(\ref{EQw5.34}),
is thermodynamically closed and Eq.~(\ref{EQw6.16}) is the traceless,
diagonally symmetric, total energy--momentum tensor and the
differential equation, Eq.~(\ref{EQw6.20}), is a tensor conservation
law.
\par
All of the quantities that constitute the total energy density,
total momentum density, and total energy--momentum tensor are
electromagnetic quantities with the caveat that the gradient of the
refractive index is small.
Although rigorous results are restricted to a limiting case, the
the real-world necessity of a non-zero gradient 
adds only a small perturbative effect.
The opposite limit of a piecewise homogeneous medium without an 
antireflection coating must be handled using Fresnel boundary conditions 
\cite{BIFresnel}.
\par
\section{Experimental Confirmation}
\par
\subsection{The Balazs thought experiment}
\par
In 1953, Balazs \cite{BIBalazs} proposed a thought experiment to resolve
the Abraham--Minkowski controversy.
The thought experiment was based on the law of conservation of momentum
and a theorem that the center of mass, including the rest mass that is
associated with the energy, moves at a uniform velocity \cite{BIBoyer}.
The total energy
\begin{equation}
E=\left ( {\bf p}\cdot{\bf p}c^2+m^2c^4 \right )^{1/2}
\label{EQt7.01}
\end{equation}
becomes the Einstein formula $E=mc^2$ for massive particles
in the limit ${\bf v}/c \rightarrow 0$.
For massless particles, like photons, Eq.~(\ref{EQt7.01}) becomes
\begin{equation}
{\bf p}=\frac{E}{c} {\bf \hat e}_k =
\frac{\hbar\omega_0}{c} {\bf \hat e}_k \, ,
\label{EQt7.02}
\end{equation}
where ${\bf \hat e}_k$ is a unit vector in the direction of motion.
Equation (\ref{EQt7.02}) defines the instantaneous momentum of a
photon between scattering events in a microscopic model of a dielectric.
The description of the macroscopic momentum of a field in terms of the
momentums of constituent photons is difficult because the effective
momentum of a photon in the direction of propagation of the macroscopic
field is different from its instantaneous momentum due to scattering.
Some sort of averaging process is required, at which point the
single photon description becomes a problem.
An additional issue with the photon description of light propagation in
a continuous dielectric is illustrated by the commingling of macroscopic
fields and the macroscopic refractive index with microscopic photon
momentum and momentum states in a description of photon recoil momentum
in a medium \cite{BICampbell}.
There are other complications, including an indefinite photon number,
that cause us to choose to choose a macroscopic classical description
for light propagation in a dielectric.
\par
As an electromagnetic field propagates from vacuum into a simple linear
dielectric, the effective velocities of photons in the field are reduced
due to scattering.
There is a corresponding increase in photon density in the dielectric.
Likewise, the classical energy density $({\bf \Pi}^2+{\bf B}^2)/2$
and the classical momentum density ${\bf B}\times{\bf \Pi}/c$ are 
enhanced by a factor of $n$ in the dielectric, compared to the vacuum.
For finite pulses in a dielectric, the enhanced energy density is 
offset by a narrowing of the pulse so that the electromagnetic energy 
\begin{equation}
U_{total}=\int_{\Sigma}\frac{1}{2}\frac {{\bf\Pi}^2+{\bf B}^2}{c}dv \, ,
\label{EQt7.03}
\end{equation}
is time independent for quasimonochromatic fields in the plane-wave
limit.
The electromagnetic energy is the total energy by virtue of being
constant in time.
Likewise, the electromagnetic momentum,
\begin{equation}
{\bf G}_{total}=\int_{\Sigma} \frac{{\bf B}\times{\bf \Pi}}{c} dv \,,
\label{EQt7.04}
\end{equation}
is time independent and is the total momentum.
The center-of-energy velocity of the field slows to
$(c/n){\bf \hat e}_k$.
\par
Invoking the Einstein mass--energy equivalence, it is argued in the
scientific literature \cite{BIBarn} that some microscopic constituents
of the dielectric must be accelerated and then decelerated by the field;
otherwise the theorem that the center of mass--energy moves at a
constant velocity is violated.
For a distribution of particles of mass $m_i$ and velocity ${\bf u}_i$,
the total momentum
\begin{equation}
{\bf P}_{total}= \sum_i m_i {\bf u}_i 
\label{EQt7.05}
\end{equation}
is the sum of the momentums of all the particles $i$ in the
distribution.
If the mass of each particle $m_i$ is constant, the statement that the
velocity of the center of mass
\begin{equation}
{\bf u}_{CM}= \frac{\sum_i m_i {\bf u}_i}{\sum_i m_i}
\label{EQt7.06}
\end{equation}
is constant is a statement of conservation of total momentum.
\par
Because of the enhanced momentum density of the field in a dielectric,
the differential of electromagnetic momentum
\begin{equation}
\delta {\bf p}=\frac{{\bf B}\times{\bf \Pi}}{c} \delta v
\label{EQt7.07}
\end{equation}
that is contained in an element of volume $\delta v$ (a ``particle''),
is a factor of $n$ greater than in the vacuum.
For a finite pulse, the narrower pulse width and enhanced momentum 
density offset allowing the electromagnetic momentum to be constant in
time as the field enters, and exits, the dielectric through the
gradient-index antireflection coating.
Consequently, there is no need to hypothesize any motion of the material
constituents of the dielectric to preserve the conservation of linear
momentum, even though the velocity of light slows to $c/n$.
\par
\subsection{The Jones--Richards experiment}
\par
One of the enduring questions of the Abraham--Minkowski controversy is
why the Minkowski momentum is so often measured experimentally while
the Abraham form of momentum is so favored in theoretical work.
We now have the tools to answer that question.
The Minkowski momentum is not measured directly, but inferred from a
measured index dependence of the optical force on a mirror placed in a
dielectric fluid \cite{BIPfei,BIAMC2,BIExp}.
The force on the mirror is
\begin{equation}
{\bf F} = \frac{d}{d\bar x_0}(2c{\bf G})
=\frac{d}{d\bar x_0}\int_V
2{\bf B}\times{\bf \Pi} \, \delta (z)dv \, ,
\label{EQt7.08}
\end{equation}
which depends on the total momentum density, Eq.~(\ref{EQw6.10}).
If we were to assume ${\bf F}=2d{\bf G}/dt$, which is the relation
between momentum and force in an otherwise empty spacetime,
then we would write 
\begin{equation}
{\bf F} = \frac{1}{c}\frac{d}{dt}\int_V
{2{\bf D}\times {\bf B}} \, \delta (z) dv \, .
\label{EQt7.09}
\end{equation}
Then one might infer from Eq.~(\ref{EQt7.09}) that the momentum density
of the field in the dielectric fluid is the Minkowski momentum density.
\par
The measured force on the mirror in the Jones--Richards
experiment \cite{BIExp}
is consistent with both Eqs.~(\ref{EQt7.08}) and Eqs.~(\ref{EQt7.09}),
depending on what theory you use to interpret the results.
Clearly an experiment that measures force, instead of directly measuring
the change in momentum in the dielectric, will not conclusively
distinguish the momentum density.
Specifically, the Jones--Richards experiment does not prove that the
Minkowski momentum density is the momentum density in the dielectric,
as has been argued, nor does it prove that the total momentum density,
Eq.~(\ref{EQw6.10}), is the momentum density in the dielectric.
However, based on the changes to continuum electrodynamics that are
necessitated by conservation of energy and momentum in the propagation
of light in a continuous medium, we can justify Eq.~(\ref{EQt7.08}),
instead of Eq.~ (\ref{EQt7.09}), as the appropriate relation between the
force on the mirror and the momentum of the field in a dielectric.
\par
\section{Conclusion}
\par
It has been said that physics is an experimental science and that
physical theory must be constructed on the solid basis of observations
and measurements.
That is certainly true for serendipitous discoveries like x-rays and 
radioactivity; But Maxwell \cite{BIMaxwell} used inductive reasoning to
modify the Amp\`ere law and construct the laws of electromagnetics two
decades before Hertz \cite{BIHertz} demonstrated the existence of
electromagnetic waves.
Later, Einstein's theory of special relativity violated the
well-established and experimentally verified law of conservation of
mass and this law was modified to become the law of
conservation of mass--energy.
Mathematics is the language of physics and there are many other examples
(quantum mechanics, nonlinear optics, high-energy particle physics,
etc.) where theory led experiments and not the other way around.
\par
In this article, we treated Maxwellian continuum electrodynamics as an
axiomatic formal theory and showed that valid theorems of the formal theory
are contradicted by conservation laws.
Axiomatic formal theory is a cornerstone of abstract mathematics and the
contradiction of valid theorems of Maxwellian continuum electrodynamics
by other fundamental laws of physics proves, unambiguously, that
electrodynamics and energy--momentum conservation laws, as currently
applied to dielectrics, are mutually inconsistent.
\par
We then established a rigorous basis for a reformulation of theoretical
continuum electrodynamics by deriving equations of motion for the
macroscopic fields from Lagrangian field theory adapted for a
dielectric-filled spacetime.
We reformulated the conservation laws, which were originally derived
for the case of an unimpeded inviscid flow of non-interacting
particles (dust, fluid, etc.) in the continuum limit in an otherwise empty
volume for the flow of a light field in a dielectric-filled volume.
In a separate publication \cite{BIPrepr}, we used coordinate
transformations between inertial reference frames in a dielectric-filled
volume to derive a theory of dielectric special relativity.
The reformulated versions of continuum electrodynamics, special
relativity, spacetime, field theory, and energy--momentum conservation
laws are mutually consistent in a dielectric-filled volume.
The Abraham--Minkowski controversy is trivially resolved because
the tensor total energy--momentum continuity theorem, the total
energy--momentum tensor, the total momentum, and the total energy are
fully electromagnetic and unique for a closed and complete system
consisting of a simple linear dielectric block draped with a
gradient-index antireflection coating that is illuminated by
quasimonochromatic light.
The newly derived theory makes a unique prediction that was shown to be
consistent with the Balazs \cite{BIBalazs} thought experiment and the
Jones--Richards experiment \cite{BIExp} and is consequently compliant
with the Scientific Method.
\par

\end{document}